\documentclass{aa} 

\usepackage{graphicx}
\usepackage{natbib}
\usepackage{capt-of}
\usepackage{textcomp}
\usepackage{epstopdf}
\usepackage{rotating}
\usepackage{siunitx}
\usepackage{multirow}
\usepackage{hhline}
\usepackage{float}
\floatstyle{plaintop}
\restylefloat{table}
\usepackage{subcaption}
\usepackage{graphicx}
\usepackage{hyperref}
\begin{document} 

\renewcommand{\thefootnote}{\alph{footnote}}
\renewcommand{\thefootnote}{\fnsymbol{footnote}}

   \title{Infrared spectra of complex organic molecules in astronomically relevant ice matrices I}
   \subtitle{Acetaldehyde, ethanol, and dimethyl ether}
\titlerunning{Infrared spectra of complex organic molecules}

   \author{J. Terwisscha van Scheltinga\thanks{Both authors contributed equally to this work} \inst{,1,2}, N.F.W. Ligterink \inst{\star ,1,2}, A.C.A. Boogert \inst{3,4}, E.F. van Dishoeck\inst{2,5}         
          \and H. Linnartz \inst{1}
          }
  \authorrunning{J. Terwisscha van Scheltinga et al.}

   \institute{
             %\inst{1} 
             Raymond and Beverly Sackler Laboratory for Astrophysics, Leiden Observatory, Leiden University, PO Box 9513, 2300 RA Leiden, The Netherlands\\    
                         \email{jeroentvs@strw.leidenuniv.nl}\\                      
             \email{linnartz@strw.leidenuniv.nl}\\ 
             \and
             %\inst{2}
             Leiden Observatory, Leiden University, PO Box 9513, 2300 RA Leiden, The Netherlands\\  
             \and
             %\inst{3}
             Universities Space Research Association, Stratospheric Observatory for Infrared Astronomy, NASA Ames Research Center, Moffett Field, California 94035, USA\\
             \and
             %\inst{4}
             Institute for Astronomy, University of Hawaii, 2680 Woodlawn Dr., Honolulu, HI 98622, USA\\
             \and
             %\inst{5}
             Max-Planck Institut für Extraterrestrische Physik (MPE), Giessenbackstr. 1, 85748 Garching, Germany\\
             }

\abstract
  % context heading (optional)
{The number of identified complex organic molecules (COMs) in inter- and circumstellar gas-phase environments is steadily increasing. Recent laboratory studies show that many such species form on icy dust grains. At present only smaller molecular species have been directly identified in space in the solid state. Accurate spectroscopic laboratory data of frozen COMs, embedded in ice matrices containing ingredients related to their formation scheme, are still largely lacking.}    
  % aims heading (mandatory) 
{This work provides infrared reference spectra of acetaldehyde (CH$_3$CHO), ethanol (CH$_3$CH$_2$OH), and dimethyl ether (CH$_3$OCH$_3$) recorded in a variety of ice environments and for astronomically relevant temperatures, as needed to guide or interpret astronomical observations, specifically for upcoming James Webb Space Telescope observations.}
  % methods heading (mandatory) 
{Fourier transform transmission spectroscopy (500-4000 cm$^{-1}$ / 20-2.5 $\mu$m, 1.0 cm$^{-1}$ resolution) was used to investigate solid acetaldehyde, ethanol and dimethyl ether, pure or 
mixed with water, CO, methanol, or CO:methanol. These species were deposited on a cryogenically cooled infrared transmissive window at 15~K. A heating ramp was applied, during which IR spectra were recorded until all ice constituents were thermally desorbed.}
  % results heading (mandatory) 
{We present a large number of reference spectra that can be compared with astronomical data. Accurate band positions and band widths are provided for the studied ice mixtures and temperatures. Special efforts have been put into those bands of each molecule that are best suited for identification. For acetaldehyde the 7.427 and 5.803 $\mu$m bands are recommended, for ethanol the 11.36 and 7.240 $\mu$m bands are good candidates, and for dimethyl ether bands at 9.141 and 8.011 $\mu$m can be used. All spectra are publicly available in the 
Leiden Database for Ice.}
  % conclusions heading (optional), leave it empty if necessary 
{}

     
   \maketitle
%
%________________________________________________________________
\section{Introduction}
\label{sec.int}

Water was the first molecule to be detected in the solid state in the interstellar medium \citep{gillettforrest1973}. Since then more than 10 other molecules have been identified in icy form (i.e. CO, CO$_{2}$, CH$_{4}$, NH$_{3}$ and CH$_{3}$OH) and it has become clear that icy dust grains play a key role in the formation of both these small molecules and more complex organic molecules (COMs), such as glycolaldehyde (HOCH$_{2}$CHO) and ethylene glycol (HOCH$_{2}$CH$_{2}$OH). The combined outcome of astronomical observations, specifically space based missions such as the $Infrared\ Space\ Observatory$ (ISO) and $Spitzer\ Space\ Telescope$ \citep{kessler1996,werner2004}, laboratory, and astrochemical modelling studies have resulted in a detailed picture of the composition and structure of ice mantles on interstellar dust grains and the chemical processes taking place \citep[see reviews by][]{gibb2000,herbstdishoeck2009,oberg2011,caselliceccarelli2012,tielens2013,boogert2015,linnartz2015,oberg2016}. It is generally accepted that interstellar ices form on the surface of dust grains in cold dark clouds through accretion in two distinct layers: a polar H$_{2}$O-rich and an apolar CO-rich layer. Water, together with NH$_{3}$, CO$_{2}$, and CH$_{4}$, forms through atom addition reactions in lower density environments \citep{hiraoka1995,miyauchi2008,oba2009,dulieu2010,hidaka2011,linnartz2011,oba2012,lamberts2013,lamberts2014,fedoseev2015a}. At later stages, when densities increase and temperatures decrease along with the ongoing cloud collapse, CO freeze-out from the gas phase occurs, forming a CO coating on top of the water rich layer \citep{tielens1991,pontoppidan2006}. Subsequent hydrogenation processes transform CO to H$_{2}$CO and H$_{2}$CO to CH$_{3}$OH \citep{watanabe2002,fuchs2009}, resulting in CO ice intimately mixed with methanol \citep{cuppen2011,penteado2015}. Radical recombination processes in various starting mixtures, triggered by energetic (i.e. UV photons or cosmic rays) or non-energetic (i.e. atom additions) were shown to provide pathways towards the formation of more complex molecules  \citep[see reviews of][]{linnartz2015,oberg2016}.

The molecules H$_{2}$O, CO, CO$_{2}$, CH$_{4}$, NH$_{3}$, and CH$_{3}$OH make up the bulk of interstellar ice \citep{ehrenfreund2000,oberg2011}, but less abundant species have been observed as well. These include species such as OCS and OCN$^{-}$ \citep{palumbo1995,vanbroekhuizen2004}. A number of COMs, such as formic acid (HCOOH), acetaldehyde (CH$_{3}$CHO), and ethanol (CH$_{3}$CH$_{2}$OH), have been tentatively detected based on spectroscopic features at 7.2 and 7.4 $\mu$m \citep{schutte1999,oberg2011}. Several other spectroscopic features, such as the 6.0 and 6.8 $\mu$m bands, remain only partly identified \citep{schutte1996,boudin1998,gibbwhittet2002,boogert2008}. Limited astronomical detection sensitivity combined with a lack of high resolution laboratory data have thus far prohibited secure solid state identifications of COMs other than methanol, but their presence in interstellar ices is generally accepted and also further supported by the recent detection of a number of COMs on comet 67P/Churyumov-Gerasimenko and in its coma \citep{goesmann2015,altwegg2017}.

\begin{figure*}
\begin{center}
\includegraphics[width=\hsize]{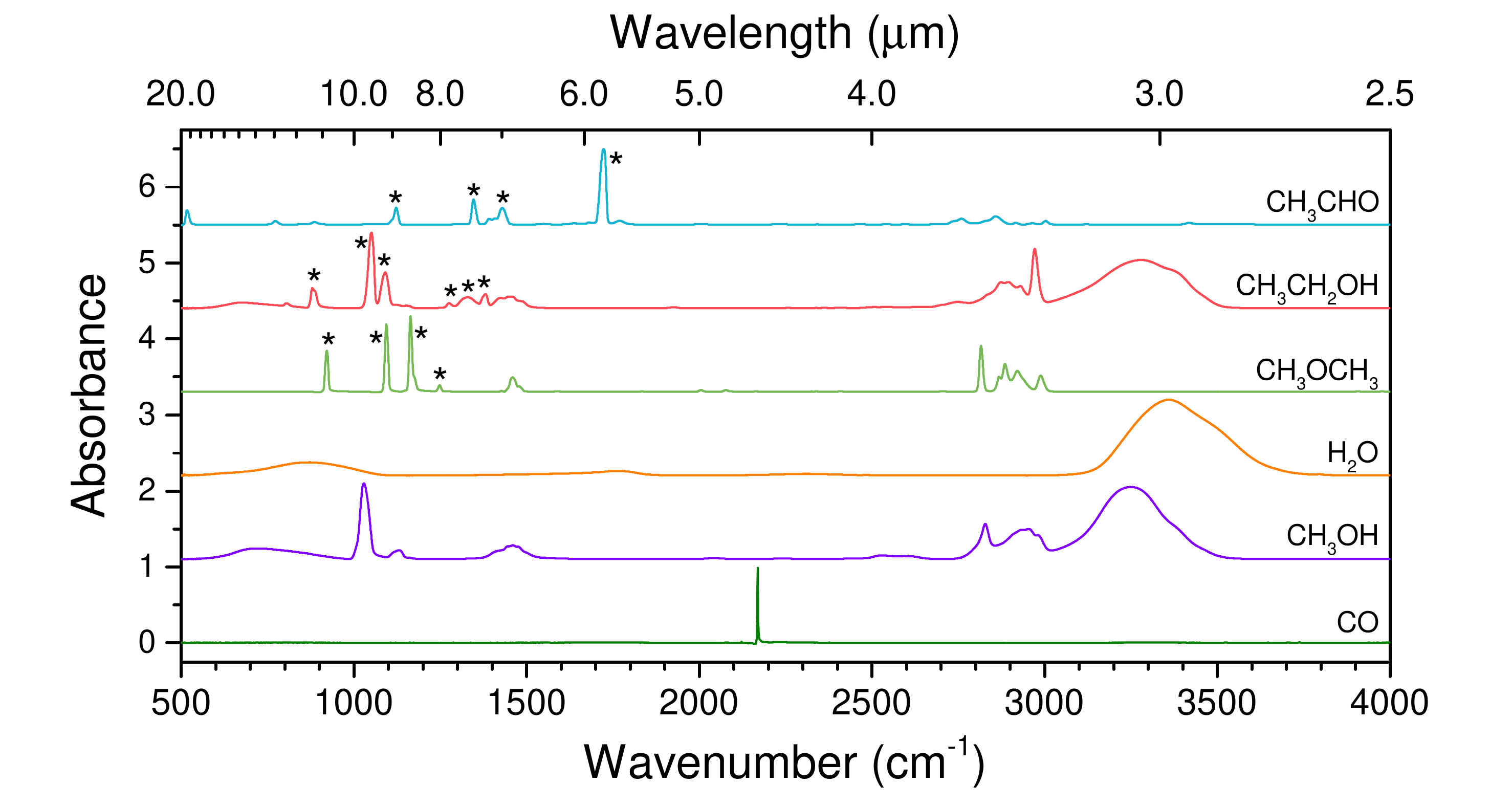}
\caption{Spectra of pure acetaldehyde (blue), ethanol (red), dimethyl ether (green), water (orange) methanol (purple), and CO (dark green) normalized to one in the range of 2.5 to 20.0 $\mu$m. The bands investigated in this work are indicated with an asterix (*).}
\label{fig.COM_comp_15K}
\end{center}
\end{figure*}

With the upcoming launch of the $James\ Webb\ Space\ Telescope$ (JWST) in 2019, new instruments such as MIRI \citep[Mid InfraRed Instrument;][]{wright2015} and NIRSpec \citep[Near InfraRed Spectrograph;][]{posselt2004} will become available to record telluric free spectra of interstellar ices at higher spectral and spatial resolution and with higher sensitivity than possible so far. This opens up new possibilities to search for and study the level of molecular complexity in interstellar ices. To aid in the search for larger molecules in the solid state, high resolution IR laboratory spectra are required. The ice matrix environment and its temperature have to be taken into account since these influence the spectral appearance of vibrational bands.

In this work we present the infrared spectra of acetaldehyde, ethanol, and dimethyl ether, respectively, CH$_{3}$CHO, CH$_{3}$CH$_{2}$OH, and CH$_{3}$OCH$_{3}$. The choice for these three species, an aldehyde, an alcohol, and an ether, is motivated by previous tentative identifications \citep{boudin1998,schutte1999,oberg2011}, their astronomical gas-phase identification and high abundance \citep[e.g.][]{turner1991,gibb2000,cazaux2003,bisschop2007b,taquet2015,muller2016}, and their common formation scheme upon UV irradiation of methanol ice \citep{oberg2009a}. Formation of these molecules is seen in energetic processing experiments of methanol ice \citep{gerakines1996,bennett2007,oberg2009a,boamah2014} and starts with cleavage of the CH$_{3}$OH bonds. This results in a reservoir of radicals that can be used for their formation as follows:
\\ \\
$^{\bullet}$CH$_{3}$ + $^{\bullet}$CHO $\rightarrow$ CH$_{3}$CHO
\\ \\
$^{\bullet}$CH$_{3}$ + $^{\bullet}$CH$_{2}$OH $\rightarrow$ CH$_{3}$CH$_{2}$OH
\\ \\
$^{\bullet}$CH$_{3}$ + $^{\bullet}$OCH$_{3}$ $\rightarrow$ CH$_{3}$OCH$_{3}$
\\ \\
Formation of dimethyl ether and ethanol has also been studied by radical recombination reactions starting from CH$_{4}$:H$_{2}$O mixtures \citep{bergantini2017}. Besides energetic radical recombination reactions, other formation pathways and links between the three molecules exist as well. For example, acetaldehyde has been proposed as a solid state precursor of ethanol. A hydrogen atom addition experiment showed that acetaldehyde can at least partially (>20\%) be transformed into ethanol \citep{bisschop2007a}. 

Acetaldehyde itself has been proposed to form as a spin-off in the well-studied CO+H $\rightarrow$ HCO $\rightarrow$ H$_2$CO $\rightarrow$ H$_3$CO $\rightarrow$ CH$_3$OH chain \citep{charnley2004}; HCO may directly interact with a C-atom, to form HCCO that upon hydrogenation yields CH$_{3}$CHO \citep{charnleyrodgers2005}. 
\\

This work presents a detailed study of the IR spectral characteristics of CH$_{3}$CHO, CH$_{3}$CH$_{2}$OH, and CH$_{3}$OCH$_{3}$ in pure form and mixed in the interstellar relevant ice matrices H$_{2}$O, CO, CH$_{3}$OH, and CO:CH$_{3}$OH. Section \ref{sec.exp} contains the experimental details and measurement protocols. The results of the measurements are presented and discussed in section \ref{sec.res}. In section \ref{sec.w33a} the astronomical relevance of the new data is illustrated. The conclusions are summarized in Sect. \ref{sec.con}. A complete overview with all data obtained in this study is available from the Appendices.

\begin{table*}
\centering
\caption[]{Selected bands of acetaldehyde, ethanol, and dimethyl ether}
\begin{tabular}{l l l l r l l}
\hline
\hline
\noalign{\smallskip}
\multicolumn{2}{c}{\multirow{2}{*}{Species}} & \multirow{2}{*}{Formula} & \multirow{2}{*}{Mode} & \multicolumn{2}{c}{Peak position*} & $A'$ \\
&       &       & & cm$^{-1}$ & $\mu$m & cm molec.$^{-1}$ \\
\hline
\\
\multirow{4}{*}{\includegraphics[height=2.0cm]{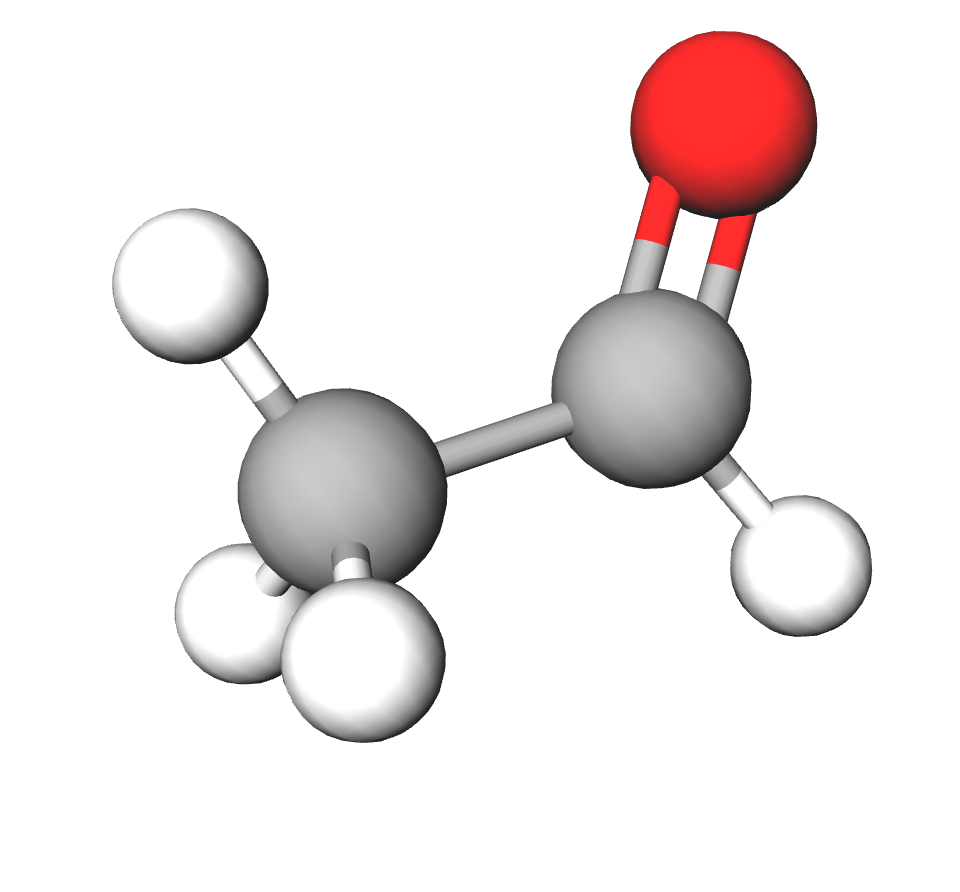}} & \multirow{5}{*}{Acetaldehyde} & \multirow{5}{*}{CH$_{3}$CHO} & CH$_{3}$ rock. + CC stretch. & 1122.3 & 8.909 \\
& & & + CCO bend. &  & \\
& & & CH$_{3}$ s-deform. + CH wag. & 1346.2 & 7.427 \\
& & & CH$_{3}$ deform. & 1429.4 & 6.995 \\
& & & CO stretch. & 1723.0 & 5.803 & 1.3$\times$10$^{-17,a}$\\
\\
\hline
\\
\multirow{6}{*}{\includegraphics[height=1.6cm]{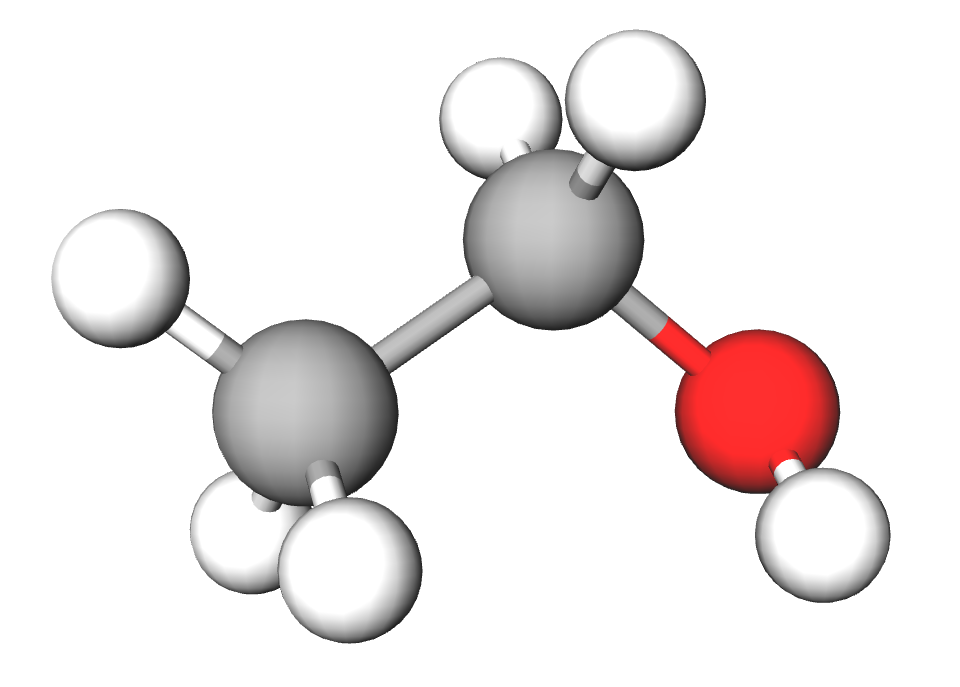}} & \multirow{6}{*}{Ethanol} & \multirow{6}{*}{CH$_{3}$CH$_{2}$OH} & CC stretch. & 879.8 & 11.36 & 3.24$\times$10$^{-18,b}$ \\
& & & CO stretch. & 1051.0 & 9.514 & 1.41$\times$10$^{-17,b}$\\
& & & CH$_{3}$ rock.  & 1090.5 & 9.170 & 7.35$\times$10$^{-18,b}$\\
& & & CH$_{2}$ tors. & 1275.2 & 7.842 \\
& & & OH deform. & 1330.2 & 7.518 \\
& & & CH$_3$ s-deform. & 1381.3 & 7.240 \\
\\
\hline
\\
\multirow{4}{*}{\includegraphics[height=2.0cm]{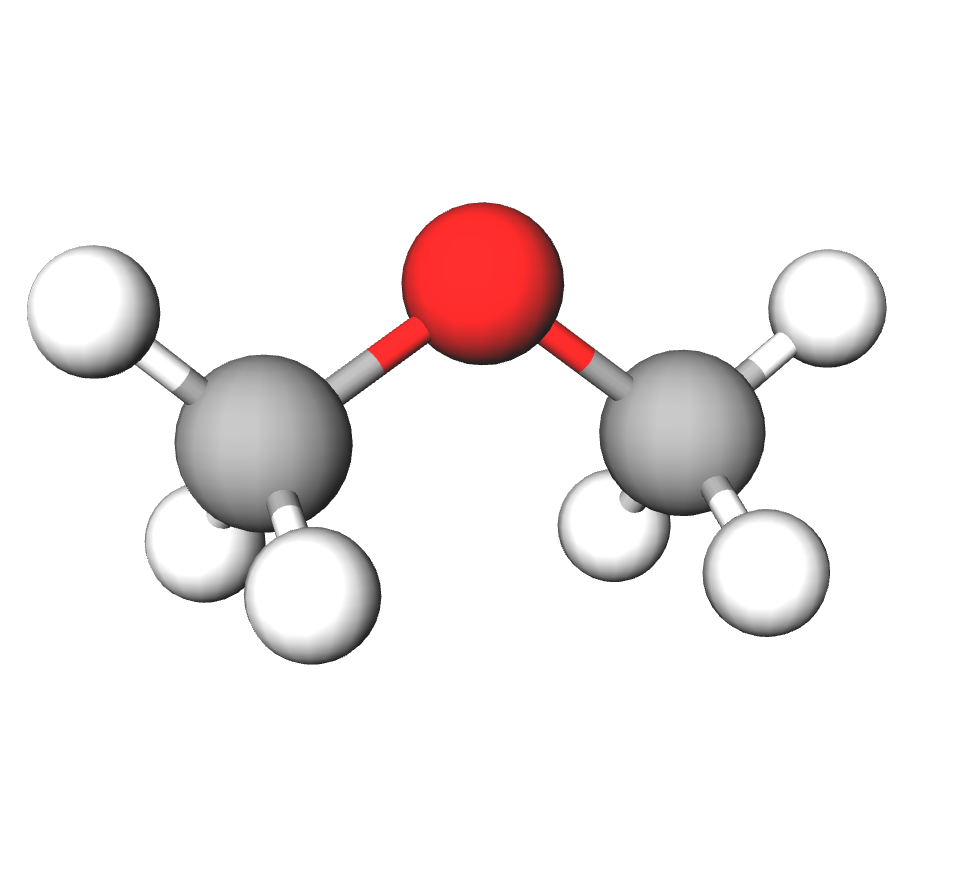}} &\multirow{4}{*}{Dimethyl ether} & \multirow{4}{*}{CH$_{3}$OCH$_{3}$} & COC stretch. & 921.3 & 10.85 & 5.0$\times$10$^{-18,c}$\\
& & & COC stretch. + CH$_{3}$ rock. & 1093.9 & 9.141 & 9.2$\times$10$^{-18,c}$\\
& & & COC stretch. + CH$_{3}$ rock. & 1163.8 & 8.592 & 9.8$\times$10$^{-18,c}$\\
& & & CH$_{3}$ rock. & 1248.2 & 8.011 \\
\\
\hline
\noalign{\smallskip}
\end{tabular}
\label{tab.selectbands}
\\
\tablefoot{*Peak position of the pure molecule at 15~K. $^{a}$\citet{schutte1999}, $^{b}$\citet{hudson2017}, $^{c}$This work. Note that throughout literature there seems to be disagreement in the assignment of certain modes, particularly for ethanol.}
\end{table*}

%__________________________________________________________________
\section{Experimental}
\label{sec.exp}

\subsection{Set-up}

The ice spectra are recorded in a high-vacuum (HV) set-up, which is described in detail by \citet{bossa2015}. A central stainless steel chamber is evacuated by a 300 l\,s$^{-1}$ turbomolecular pump, backed by a double stage rotary vane pump (8 m$^3$ hr$^{-1}$). This allows a base pressure of $\sim$10$^{-7}$ mbar at room temperature. The pressure is monitored by an Agilent FRG-720 full range gauge. Ices are grown on an infrared transmissive ZnSe window that is cryogenically cooled to a lowest temperature of 12 K by a closed cycle helium cryostat. The temperature of the window is monitored by a LakeShore 330 temperature controller, which regulates a feedback loop between a resistive heating wire and a silicon diode 
temperature sensor. An absolute temperature accuracy of $\pm$2 K and a relative accuracy of $\pm$1 K is acquired with this diode. The IR beam of a Fourier Transform InfraRed Spectrometer (FTIRS; Varian 670-IR) is aligned through the window to obtain IR spectra of the samples. The spectrometer covers a range of 4000 to 500 cm$^{-1}$ (2.5--20 $\mu$m) at spectral resolutions as high as 0.1 cm$^{-1}$. 
Samples are externally prepared in a 2L glass bulb using a separate multi-line gas mixing system. The gas mixing line is turbomolecularly pumped to pressures < 1$\times$10$^{-4}$ mbar. Gas mixtures are made by sequential addition of its components. Two gas independent gauges, covering various pressure ranges ensure that accurate mixing ratios are obtained with a maximum error of $<$10\%. The liquids and gases used in these experiments are acetaldehyde (Sigma-Aldrich, 99.5\%), ethanol (Biosolve, 99.9\%), dimethyl ether (Sigma-Aldrich, 99.9\%), water (Milli-Q, Type I), carbon monoxide (Linde gas, 99.997\%), and methanol (Sigma-Aldrich, 99.9\%). Liquid samples are purified with freeze-pump-thaw cycles before use. 

\subsection{Measurement protocol}

Pure or premixed gases are background deposited onto the 15 K cold sample via an inlet valve. A standard pressure of 20 mbar in the glass bulb is used to prevent a decreasing inlet pressure gradient during deposition. Bi-mixed gases are prepared in a 1:20 ratio and tri-mixed gases in a 1:20:20 ratio, where the smallest fraction is the COM under investigation. These dilution factors ensure that the COM mainly interacts with the surrounding matrix, resulting in matrix shifted IR vibrational bands. Ices are grown at 15~K to a column density of $\sim$4500 ML (1 monolayer is 1$\times$10$^{15}$ molecules cm$^{-2}$) on the window. This coverage ensures that any influence of background contamination, mainly water depositing at a rate of less than 30 ML hr$^{-1}$, can be neglected. During deposition, IR spectra are recorded at 1 cm$^{-1}$ resolution (0.5 cm$^{-1}$ step size) and averaged over 61 scans (equals 2 minutes) to trace the ice growth and determine when the ice is $\sim$4500 ML thick. From the integration of the IR band absorption, the column density of the species $N_{\rm species}$ is determined according to 
\begin{equation}
\label{eq.coldens}
N_{\rm species} = {\rm ln(10)}\frac{\int_{\rm band}log_{\rm 10}\left(\frac{I_{\rm 0}(\tilde{\nu})}{I(\tilde{\nu})}\right) d\tilde{\nu}}{A'_{\rm }},
\end{equation} 

where $\int_{\rm band}log_{\rm 10}\left(\frac{I_{\rm 0}(\tilde{\nu})}{I(\tilde{\nu})}\right) d\tilde{\nu}$ is the integrated absorbance of the band and $I_{\rm 0}(\tilde{\nu})$ and $I(\tilde{\nu})$ are the flux received and transmitted by the sample, respectively, and $A'_{\rm}$ is the apparent band strength in cm molecule$^{-1}$. It is important to realize that strongly absorbing bands may get saturated at high coverages, resulting in unreliable column density measurements. In the experiments conducted, the CO band at 2135 cm$^{-1}$ reaches saturation at high coverage, as do certain bands of pure acetaldehyde and dimethyl ether. For these species, bands with a lower band strength or isotopologues can be used. The measured column densities give an indication whether the mixed ice composition still matches the gas-phase mixing ratio and whether the COMs are sufficiently diluted in the matrix. Small variations in the composition of the gas mixture and matrix interactions complicate accurate ice mixing ratio determinations. The apparent band strengths are listed in Table~\ref{tab.selectbands} and taken from literature for acetaldehyde and ethanol. For the dimethyl ether bands at 923, 1095, and 1164 cm$^{-1}$ the band strength value is approximated from a methane:dimethyl ether mixture, prepared at a one to one ratio in the gas phase. Assuming this ratio is maintained in the ice and matrix interactions are negligible, the apparent band strength is determined from a comparison with the methane 1302 cm$^{-1}$ band area and its known apparent band strength of 8.0$\times$10$^{-18}$ cm molecule$^{-1}$ \citep{bouilloud2015}.

After deposition the sample is linearly heated at a rate of 25 K hr$^{-1}$, until it is fully desorbed from the window. The low temperature ramp ensures that the ice has sufficient time to undergo structural changes, particularly from the amorphous to the crystalline phase. During heating IR spectra are continuously recorded and averaged over 256 scans to trace spectral changes versus temperature. 

\subsection{Analysis protocol}
\label{subsec.analysis}

Owing to the very large amount of spectra that are recorded during the experiments, we only present samples of representative IR spectra for temperatures at which significant spectral changes occur. These spectra are baseline subtracted and the peak position and band width at full width at half maximum (FWHM) are determined for selected spectral features. When the band of a COM overlaps with a spectral feature of a matrix molecule, also the matrix feature is subtracted where possible. In the case of band splitting, the least intense component is only taken into account when its peak position is clearly distinguishable. In a few cases splitted peaks rival in intensity and are heavily overlapping and it is not possible to fit a FWHM for the individual components. Here the FWHM of the combined peaks is determined. Peaks are selected for analysis mainly based on their intensity and potential as an ice tracer, i.e. selecting wavelengths for which no strong overlap with known interstellar features exist. 

Identification of vibrational modes of the three species studied here is realized by comparison with available spectra from liquid and solid state literature \citep{plyler1952,evansbernstein1956,barnes1970,allan1971,hollenstein1971,mikawa1971,coussan1998}. Optical effects such as longitudinal optical - transverse optical (LO-TO) splitting and particle shape effects are not explicitly taken into account. Since spectra are recorded at normal incidence with unpolarized light, only the TO modes are recorded. However, certain combinations of polarized light and angles of incidence can result in the LO phonon mode showing up \citep{baratta2000,palumbo2006}. Also particle shape effects can shift transition bands with respect to recorded laboratory spectra \citep{barattapalumbo1998}. Such effects affect only the spectra of more abundant species, such as CO or CO$_{2}$, and are not considered to be relevant for COMs.

%__________________________________________________________________
\section{Results and discussion}
\label{sec.res}

\begin{figure*}[ht]
\includegraphics[height=\hsize]{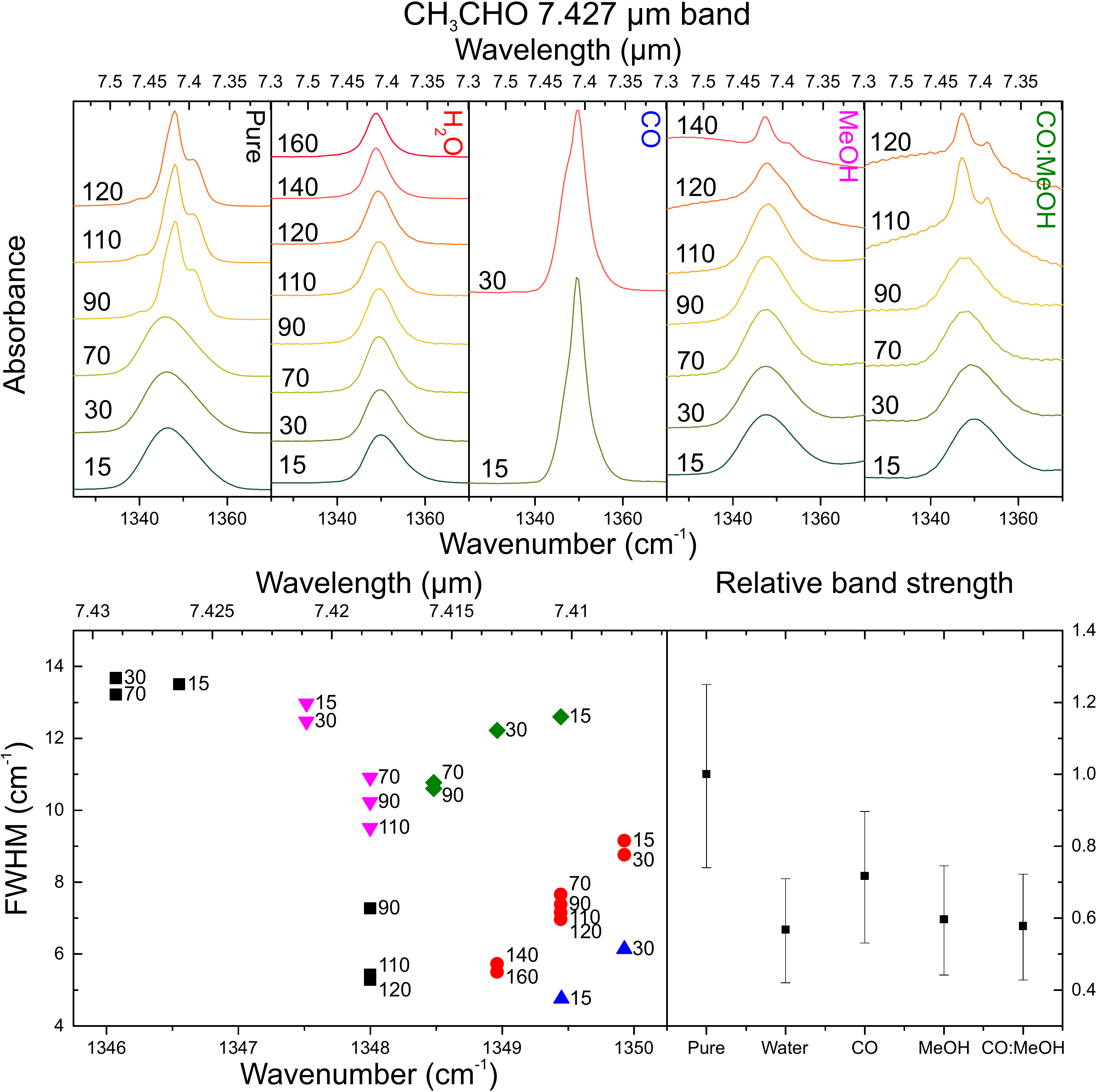}
\caption{Top: from left to right the acetaldehyde 7.427 $\mu$m band pure (black) and in water (red), CO (blue), methanol (purple), and CO:CH$_{3}$OH (green) at various temperatures. Bottom left: peak position vs. FWHM plot, using the same colour coding. Bottom right: the relative band strength for the 7.427 $\mu$m band at 15~K in various matrices.}
\label{fig.A1350}
\end{figure*}

\begin{table*}
\centering
\caption[]{Peak position and FWHM of the acetaldehyde CH$_{3}$ s-deformation + CH wagging mode at 15~K in various matrices.}
\begin{tabular}{| l | c | l l | l l | l l |}
\hline
        \multirow{2}{*}{Mixture} & Temperature &  \multicolumn{2}{c |}{$\lambda_{\rm{peak,-baseline}}$} &  \multicolumn{2}{c |}{$\lambda_{\rm{peak,-matrix}}$} & \multicolumn{2}{c |}{FWHM} \\ 
         & (K) & (cm$^{-1}$) & ($\mu$m) & (cm$^{-1}$) & ($\mu$m) & (cm$^{-1}$) &  ($\mu$m) \\ 
        \hline
CH$_3$CHO                               &       \multirow{5}{*}{15}     &       1346.6  &       7.4264  &       --      &       --      &       13.5    &       0.0744* \\
CH$_3$CHO :H$_2$O               &       &       1349.9  &       7.4078  &       1349.9  &       7.4078  &       \phantom{1}9.2  &       0.0502  \\      
CH$_3$CHO :CO                   &       &       1349.4  &       7.4104  &       --      &       --      &       \phantom{1}4.8  &       0.0262  \\      
CH$_3$CHO :CH$_3$OH             &       &       1347.5  &       7.4211  &       --      &       --      &       13.0    &       0.0714  \\      
CH$_3$CHO :CO:CH$_3$OH  &       &       1349.4  &       7.4105  &       --      &       --      &       12.6    &       0.0691  \\      
        \hline
        \noalign{\smallskip}
\end{tabular}
\tablefoot{Excerpt from Table \ref{tab.acetal_cc_stretch}. *FWHM result of two or more blended peaks.}
\label{tab.example}
\end{table*}

In this section selected results of the acetaldehyde, ethanol, and dimethyl ether experiments are presented. These are representative for the much larger data set given in the Appendix. All the selected spectra used in this work are publicly available from the Leiden Database for Ice (http://icedb.strw.leidenuniv.nl), spectra recorded for other temperatures are available on request. Figure \ref{fig.COM_comp_15K} shows the IR spectra of pure acetaldehyde, ethanol, and dimethyl ether ice at 15~K; the bands that are fully analysed are indicated with an asterix (*) and spectra of pure water, CO, and methanol ice. Figures of the spectra of COMs mixed in water, CO, methanol, and CO:methanol at 15~K are shown in Appendix \ref{ap.spectra}. In Table \ref{tab.selectbands}, these selected bands are listed together with their peak positions and, if available, apparent band strength in pure ices at 15~K. Appendix \ref{ap.tables} presents the results of the analysis of the selected bands, listing peak positions, FWHMs, and integrated absorbance ratios at various temperatures and for different ice matrices. A representative example of the tables listed in the Appendix is shown in Table \ref{tab.example} for the acetaldehyde CH$_{3}$ s-deformation + CH wagging mode at 1346.6 cm$^{-1}$ at 15 K.

\begin{figure*}[ht]
\includegraphics[height=\hsize]{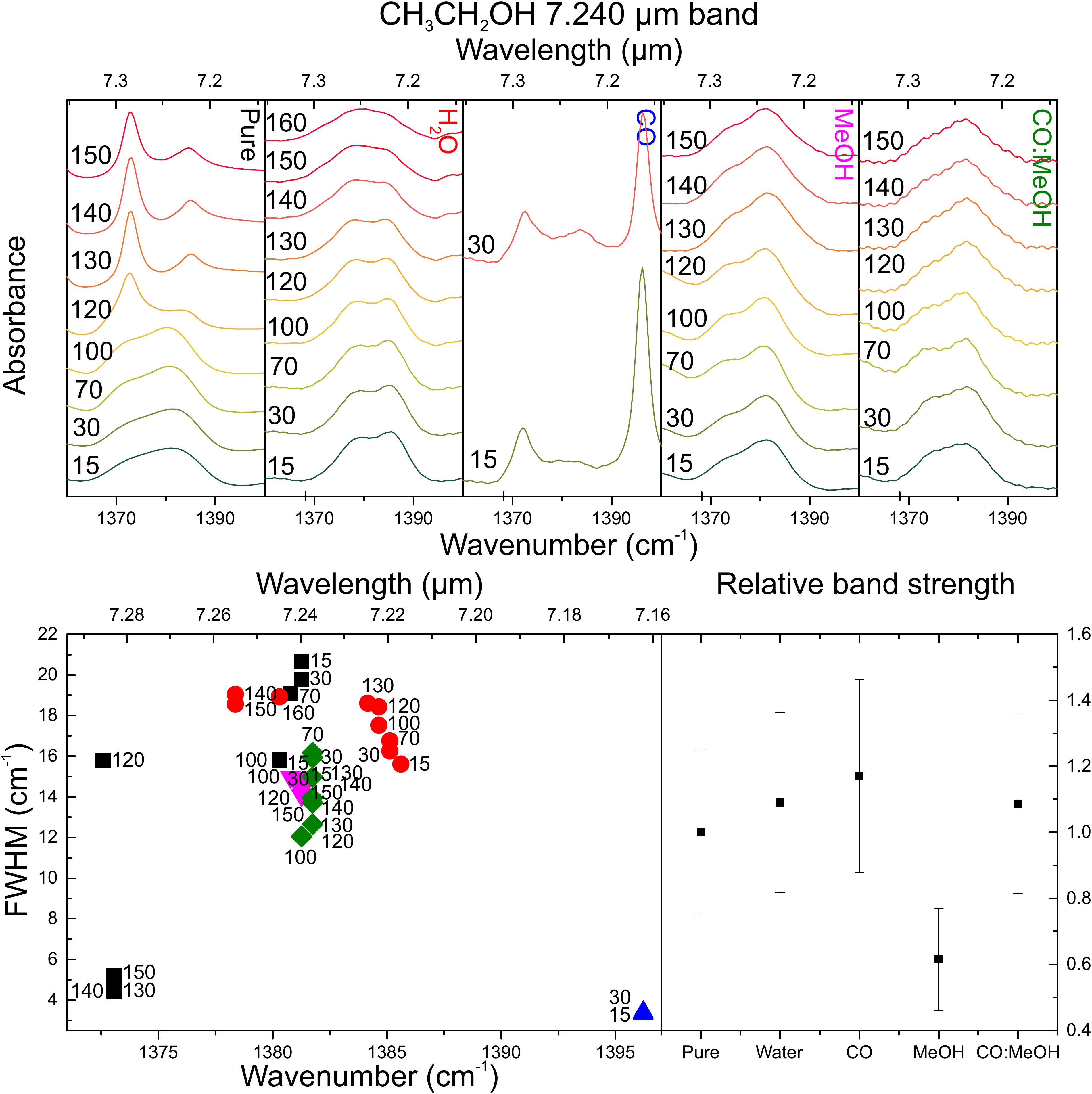}
\caption{Top: from left to right the ethanol 7.240 $\mu$m band pure (black) and in water (red), CO (blue), methanol (purple), and CO:CH$_{3}$OH (green) at various temperatures. Bottom left: peak position vs. FWHM plot, using the same colour coding. Bottom right: the relative band strength for the 7.240 $\mu$m band at 15~K in various matrices.}
\label{fig.E1381}
\end{figure*}

\begin{figure*}
\includegraphics[height=\hsize]{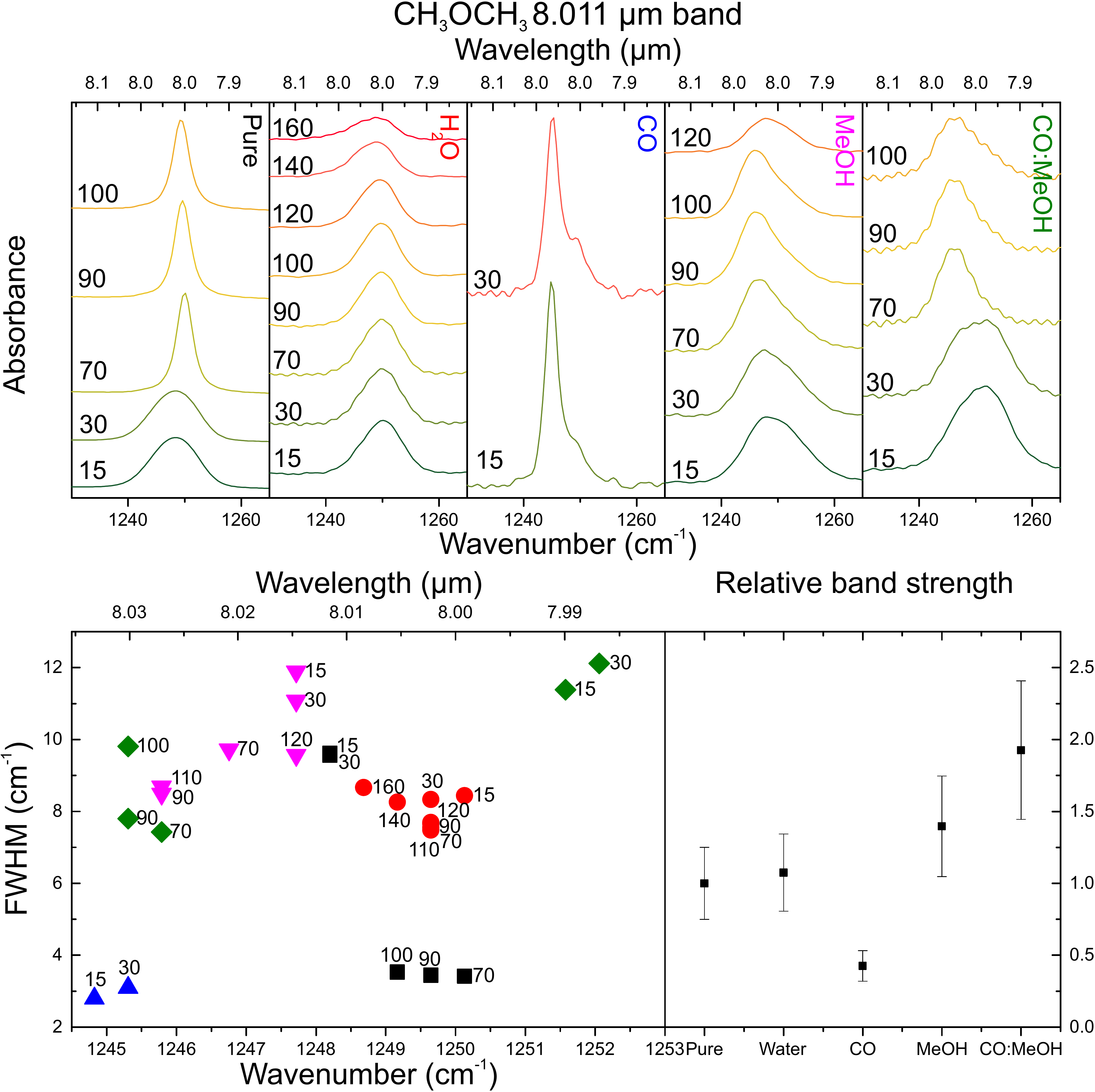}
\caption{Top: from left to right the dimethyl ether 8.011 $\mu$m band pure (black) and in water (red), CO (blue), methanol (purple), and CO:CH$_{3}$OH (green) at various temperatures. Bottom left: peak position vs. FWHM plot, using the same colour coding. Bottom right: the relative band strength for the 8.011 $\mu$m band at 15~K in various matrices.}
\label{fig.D1248}
\end{figure*}

For easier interpretation the results are represented in a number of plots; see Figs. \ref{fig.A1350}, \ref{fig.E1381}, and \ref{fig.D1248} for examples. Each plot covers the data of one band. In all plots, the top panels show spectroscopic changes of the band under thermal processing in pure and mixed ices. The bottom left panels plot peak position versus FWHM, showing trends in the band. The bottom right panels give an indication of how the band strengths change relative to each other in various matrices. Assuming that the ice column density is roughly the same for each experiment and that the gas mixing ratio is close to the ice mixing ratio, the mixed ices are corrected for their dilution factor. Owing to various uncertainties in this method, this results in relatively large error bars for the band strengths. This is unfortunate as this would allow us to interpret the spectroscopic identifications - the primary goal of this work - also in terms of accurate column densities. The remaining figures of other bands can be found in Appendix \ref{ap.visual}.

A few general statements can be made. Most peaks display peak narrowing under thermal processing, which is due to the ice changing to a crystalline phase with increasing temperature. Mixed ice in CO and CO:CH$_{3}$OH are exceptions due to the volatility of CO and its removal from the ice at relatively low temperatures. Above 30~K, the desorption temperature of CO \citep{oberg2005}, these ices are often seen to display peak broadening. 

Peak splitting, especially at high temperatures is another effect that is generally seen. This can be caused by two or more modes contributing to a single feature at low temperatures and becoming visible as the peaks begin to narrow at higher temperatures. Alternatively, the matrix can play a role and a peak is split owing to different interactions of a functional group with its surroundings. For example, an ice can segregate under thermal processing and have part of the COM still intimately mixed with the matrix molecule, while another part is forming COM clusters. Segregation is an effect most clearly seen in the COM:CO ice mixtures.

Integrated absorbance ratios are given for the bands under investigation in each ice mixture. These ratios can provide a tool to estimate the likelihood of observing other bands upon detection of a specific transition. They can also be used as conversion factors to determine band strengths from known band strengths. The bands are normalized on the band with highest integrated absorbance at 15~K, unless this band is suspected to be in saturation or when the data set is incomplete over the investigated temperature range. 

\subsection{Acetaldehyde}

Acetaldehyde hosts four significant features in the 5.5 - 12.5 $\mu$m region (see Fig. \ref{fig.COM_comp_15K}). Some smaller features are also visible, such as the CC stretching + CH$_{3}$ rocking mode close to 11 $\mu$m, however, its intensity is very small compared to the other bands. Two characteristic vibrational modes of acetaldehyde at 6.995 and 8.909 $\mu$m coincide with methanol CH$_{3}$ rocking and deformation modes and are likely obscured in interstellar spectra. A solid state identification of acetaldehyde based on these vibrational modes is unlikely. The CO stretching mode is the most prominent band in this spectrum. However, its location at 5.8 $\mu$m coincides with the CO stretching mode of many other molecules, such as formaldehyde (H$_{2}$CO), formic acid (HCOOH), or formamide (NH$_{2}$CHO), which are expected to be present in interstellar ice, making it likely that this band is blended. The fourth band is the CH$_{3}$ s-deformation + CH wagging mode around 7.427 $\mu$m, which is found to have no substantial overlap with abundant bulk interstellar ice components and therefore is most suited for a successful solid state identification of this molecule. 

Figure \ref{fig.A1350} shows the results of the analysed data of the CH$_{3}$ s-deformation + CH wagging band. Under thermal processing the band widths are generally seen to decrease; this is caused by crystallization in the ice. Peak positions shift as well, with clear blue shifting trends visible for the CO:CH$_{3}$OH and water mixtures. In the case of the CO:CH$_{3}$OH mixture this is likely because of the loss of CO from the matrix, while for the water mixture the interaction between acetaldehyde and crystalline water is more likely the cause. In some cases, at high temperature CH$_{3}$CHO undergoes peak splitting, making identification through FWHM challenging. However, this can also be used as a tool to determine the ice temperature. The comparison of peak position makes it in general easy to distinguish between
pure acetaldehyde, acetaldehyde mixed in CH$_{3}$OH, and CO:CH$_{3}$OH, acetaldehyde mixed in CO, and
acetaldehyde mixed in water. The 7.427 $\mu$m band shows a substantial decrease in band strength by about 40\% when acetaldehyde is surrounded by matrix molecules.

The acetaldehyde CO stretching band underlines the above findings, given it is clearly observed (see Fig. \ref{fig.A1720}). Especially at low ice temperatures of 15 and 30~K clear peak shifts are visible between the CO:CH$_{3}$OH matrix at 5.84 $\mu$m, the water matrix at 5.825 $\mu$m, and the pure matrix, or in a CH$_{3}$OH matrix at around 5.805 $\mu$m.

\subsection{Ethanol}

The spectrum of pure ethanol in Fig. \ref{fig.COM_comp_15K} shows a strong CC stretching band at 11.36 $\mu$m, CO stretching mode at 9.514 $\mu$m, and CH$_{3}$ rocking mode at 9.170 $\mu$m. A number of weaker modes are seen between 6.5 and 8.5 $\mu$m: specifically the CH$_{2}$ torsion mode at 7.842 $\mu$m, the OH deformation mode at 7.518 $\mu$m, and the CH$_{3}$ symmetric deformation mode at 7.240 $\mu$m. Overlap with spectral features of bulk interstellar ice species such as water and methanol is an issue for the three strongest bands, coinciding with either the water libration mode or CO stretching and CH$_{3}$ rocking modes of methanol. Also the prominent broad silicate feature is present at $\sim$9.7 $\mu$m. Although the other ethanol modes are substantially weaker, they fall within a spectral region that is generally clean of strong transitions. 

The ethanol 7.240 $\mu$m band is a possible candidate for identification. Figure \ref{fig.E1381} shows the data of this band. Ethanol mixed in water can be distinguished from other features by a $\sim$ 3 cm$^{-1}$ peak shift from other mixtures. In general it is found that the ethanol:water mixture is relatively easy to distinguish, but the other mixtures display much overlap in peak position and FWHM. The CH$_{2}$ torsion, OH deformation mode, and CH$_{3}$ symmetric deformation mode are hard to identify in the ethanol:CO mixture owing to the appearance of many more modes. Band areas and relative band strengths of these modes are therefore not considered. The band strength is seen to vary substantially for the various bands, but does not show a clear trend.

\subsection{Dimethyl ether}

Three strong bands of dimethyl ether are found at 10.85, 9.141, and 8.592 $\mu$m for the COC stretching and two COC stretching + CH$_{3}$ rocking modes, respectively. A much weaker CH$_{3}$ rocking mode is found at 8.011 $\mu$m. The first two overlap with known interstellar ice features of methanol, water, and silicates and are therefore less suited for an identification, while the third likely falls in the wing of such features and may still be used. Even though it is a weak mode, the 8.011 $\mu$m band falls in a relatively empty region of interstellar ice spectra. This feature could therefore be most suited for a dimethyl ether identification; see Fig. \ref{fig.D1248}.

For the 8.011 $\mu$m band clear differences are seen depending on the matrix. The spectra of pure and methanol mixture are distinguishable from those of the water and CO:CH$_{3}$OH mixtures by a $\sim$ 2 cm$^{-1}$ peak shift of the low temperature spectra at 15 and 30~K. In water this band displays a narrower peak compared to the other ices. The other bands also show many clear differences in peak position and FWHM between the various ice mixtures. A characteristic peak splitting structure at low temperatures is seen for the 10.85 $\mu$m band when mixed in water, methanol, or CO:CH$_{3}$OH. Interestingly, the relative band strength shows a substantial increase in the CH$_{3}$OH and CO:CH$_{3}$OH mixtures for the 8.011 $\mu$m band. Other modes do not show such clear differences. Also  it is interesting to note the fact that the COC stretching mode has the largest band area when mixed in water, while in the other mixtures this is always the CH$_{3}$ rocking mode at 8.592 $\mu$m (see Appendix~\ref{ap.dimethylether_area}).

%__________________________________________________________________
\section{COM ice features in W33A}
\label{sec.w33a}

\begin{figure}
\begin{center}
\includegraphics[width=\hsize]{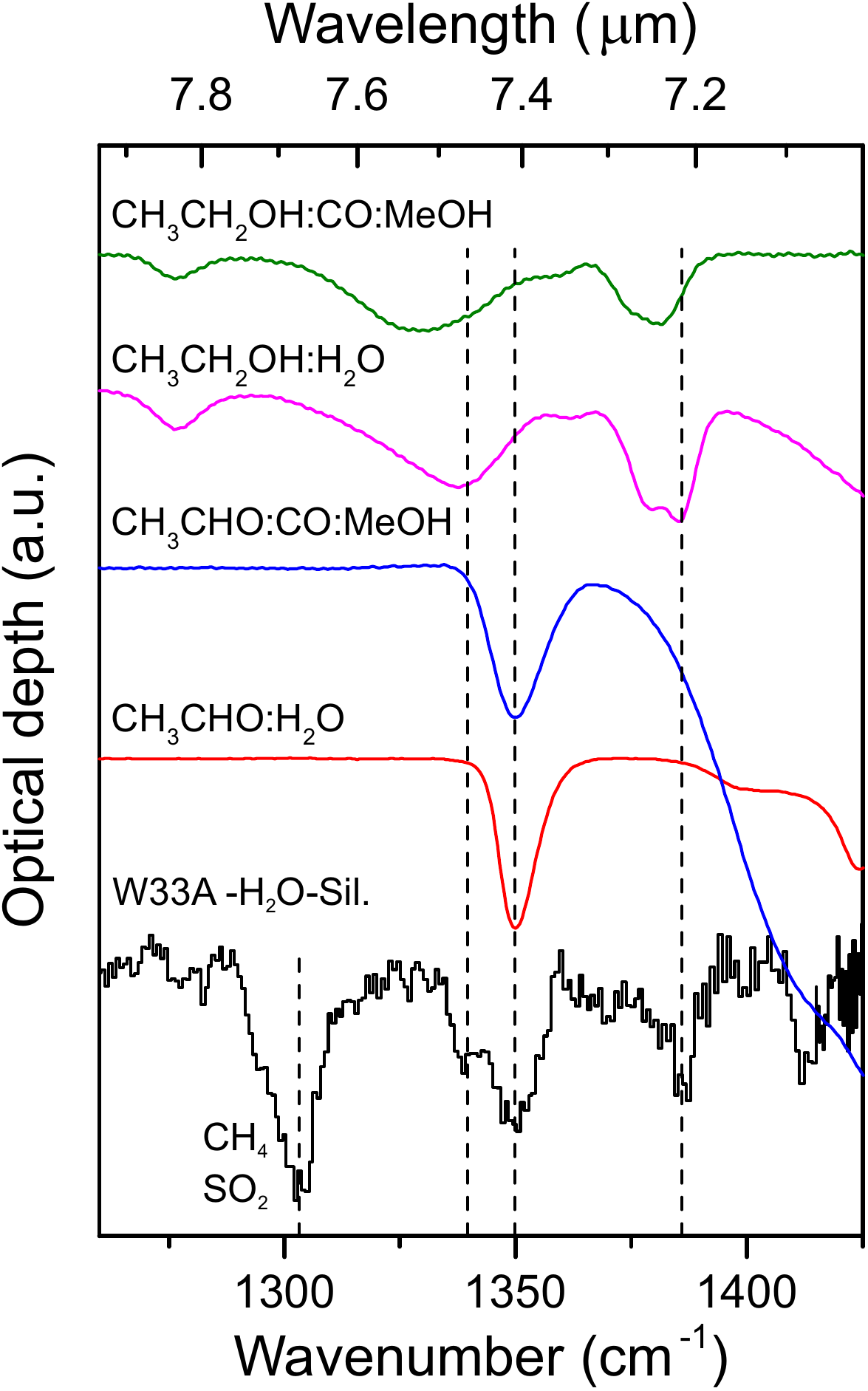}
\caption{Continuum and water and silicate subtracted spectrum of W33A plotted together with ice spectra of ethanol and acetaldehyde at 15~K, mixed in CO:CH$_{3}$OH and H$_{2}$O. Features in the W33A spectrum are indicated with dashed lines at 7.22, 7.40, and 7.47 $\mu$m. The large spectral feature at 7.67 $\mu$m is due to CH$_{4}$ and SO$_{2}$.}
\label{fig.w33a}
\end{center}
\end{figure}

Our extensive measurements of frozen COMs are needed in the analysis of the many spectra of dense clouds, embedded protostars, and inclined protoplanetary disks that will be obtained with the upcoming $JWST$ mission at high sensitivity and medium spectral resolution ($R$ of up to 3,500). Here, we demonstrate their use by a reanalysis of a spectrum of the massive protostar W 33A obtained with the Infrared Space Observatory's Short Wavelength Spectrometer (Astronomical Observation Template 1; $R$ = 800). This is one of the few
sources for which a high quality mid-IR spectrum is available \citep{gibb2000}. In the 7 to 8 $\mu$m region three prominent features at 7.25, 7.41, and 7.67 $\mu$m have been described previously in the literature. The 7.25 $\mu$m feature has been attributed to both CH$_{3}$CH$_{2}$OH and HCOOH \citep{schutte1999,oberg2011}, the 7.41 $\mu$m feature has been attributed to HCOO$^{-}$ and  CH$_{3}$CHO \citep{schutte1999}, and the 7.67 $\mu$m band has been identified as solid methane with potentially contributions of SO$_{2}$ \citep{boogert1996}. 

In this work we make use of the water and silicate subtracted spectrum of W33A, shown in Fig. \ref{fig.w33a} with a straight line local continuum subtraction. The aforementioned features are visible, although the 7.41$\mu$m feature seems to have two contributions at 7.47 and 7.40 $\mu$m and the 7.25 $\mu$m feature is found at 7.22 $\mu$m. The spectra of ethanol and acetaldehyde mixed in CO:CH$_{3}$OH and H$_{2}$O are plotted in the same figure. The peak position of the 7.40 $\mu$m feature can be reproduced well by the acetaldehyde CH$_{3}$ s-deformation mode in both mixtures. However, the band in CO:methanol mixture seems to be too broad to justly reproduce the W33A 7.40 $\mu$m feature and also this band covers the 7.47 $\mu$m feature next to it. The other two features at 7.22 and 7.47 $\mu$m could be the result of the CH$_{3}$ s-deformation and OH deformation modes of ethanol. Particularly, the CH$_{3}$CH$_{2}$OH:H$_{2}$O mixture coincides with the peak locations of the 7.22 and 7.47 $\mu$m features in the W33A spectrum. While the identification of acetaldehyde and ethanol are plausible, detection of additional features would strengthen the assignment. We checked and found that none of the other CH$_{3}$CHO and CH$_{3}$CH$_{2}$OH bands have an anti-coincidence with the W33A spectrum. 

Upper limits to the ice column densities of ethanol and acetaldehyde can be given based on the integrated optical depth of their potential features. \citet{schutte1999} give integrated $\tau$ values of 2.0$\pm$0.3 and 1.6$\pm$0.5 cm$^{-1}$, respectively. Band strength values of ethanol and acetaldehyde are taken from the literature and used to calculate the column densities of the two features. The ethanol band strength of the CO stretch mode at 9.514 $\mu$m has been determined to be 1.41$\times$10$^{-17}$ cm molecule$^{-1}$ by \citep{hudson2017}. Using the integrated absorbance ratio CH$_{3}$ s-def. / CO str. = 0.20 at 15~K from Table \ref{tab.etoh_pure_area}, the band strength of the CH$_{3}$ stretch mode is determined to be 2.8$\times$10$^{-18}$ cm molecule$^{-1}$. The effect of the matrix on the relative band strength is small for both the ethanol CO stretch and CH$_{3}$ s-deformation modes, as can be seen from Figs. \ref{fig.E1381} and \ref{fig.E1046}, and therefore assumed to be negligible. Assuming the entire 7.22 $\mu$m feature is caused by ethanol, this results in a column density of 7.1$\pm$0.2 $\times$10$^{17}$ cm$^{-2}$.

In \citet{schutte1999}, the acetaldehyde band strength is given as 1.3$\times$10$^{-17}$ cm molecule$^{-1}$ for the CO stretch mode based on data from \citet{wexler1967}. The integrated absorbance ratio of CO stretching / CH$_{3}$ s-deforming + CH wagging = 4.32 in pure acetaldehyde at 15~K in laboratory experiments. As the CO stretching mode is likely saturated, the ratio may thus be higher. Using this ratio, the band strength of the CH$_{3}$ s-deformation mode is found to be 3.0$\times$10$^{-18}$ cm molecule$^{-1}$. As can be seen in Fig. \ref{fig.A1350}, the relative band strength of this mode decreases substantially in mixtures by about 40\%. The band strength of the CH$_{3}$ s-deformation mode in mixed ices is thus 1.8$\times$10$^{-18}$ cm molecule$^{-1}$. If the entire 7.40 $\mu$m feature is attributed to CH$_{3}$CHO, the resulting column density is 8.9$\pm$3 $\times$10$^{17}$ cm$^{-2}$.

In all likelihood the 7.22 and 7.40 $\mu$m features contain contributions of other molecules, mainly HCOOH and HCOO$^{-}$ and the reported values should therefore be seen as upper limits. Using solid water and methanol column densities of 3.8$\times$10$^{19}$ and 1.7$\times$10$^{18}$ cm$^{-2}$, respectively, towards W33A \citep{dartois1999,keane2001}, the upper limit abundance ratios of ethanol and acetaldehyde can be determined. The abundance ratio $N$(COM)/$N$(H$_{2}$O) is found to be $\leq$1.9\% and $\leq$2.3\%, while $N$(COM)/$N$(CH$_{3}$OH) is $\leq$42\% and $\leq$52\% for ethanol and acetaldehyde, respectively. The abundances with respect to water are in good agreement with previously reported values of $\leq$4\% and $\leq$3.6\% for ethanol and acetaldehyde, respectively \citep{boudin1998,schutte1999}. 

The $N$(COM)/$N$(CH$_{3}$OH) upper limit ice abundance can be compared with known gas-phase abundances towards W33A. These are given as $N$(CH$_{3}$CH$_{2}$OH)/$N$(CH$_{3}$OH) = 2.4\% and $N$(CH$_{3}$CHO)/$N$(CH$_{3}$OH) < 0.2\% \citep{bisschop2007b} and are substantially lower than the ice upper limits. Interferometric observations with the Atacama Large Millimeter/submillimeter Array are needed to spatially resolve these molecules and determine more accurate abundances. Beside being upper limits, this difference may be linked to the process that transfers solid state species into the gas phase, causing molecules to fragment, or to other destruction of species in the gas phase. An overview of the COM abundances in ice and in the gas phase towards W33A is given in Table \ref{tab.abun}.

\begin{table}
\centering
\caption[]{Ice upper limits and gas-phase abundances of ethanol and acetaldehyde towards W33A. Abundances given in \%.}
\begin{tabular}{l l l l}
\hline
\hline
\noalign{\smallskip}
Species & \multicolumn{2}{c}{Ice} & Gas phase$^{c}$ \\
 & /$N$(H$_{2}$O)$^{a}$ & /$N$(CH$_{3}$OH)$^{b}$ & /$N$(CH$_{3}$OH) \\
\noalign{\smallskip}
\hline
\noalign{\smallskip}
CH$_{3}$CH$_{2}$OH & $\leq$1.9 & $\leq$42 & 2.4\\
CH$_{3}$CHO & $\leq$2.3 & $\leq$52 & $\le$0.2 \\
\hline
\noalign{\smallskip}
\end{tabular}
\label{tab.abun}
\\
\tablefoot{$^{a}$\citet{keane2001}; $^{b}$\citet{dartois1999}; $^{c}$\citet{bisschop2007b}}
\end{table}

The spectroscopic data presented in this paper, combined with the improvements in terms of sensitivity and resolution of JWST, will aid in confirming these detections and distinguish other potential contributors to these features. More observations, particularly towards low-mass sources, will give additional information about the carriers of these features.

%__________________________________________________________________
\section{Conclusions}
\label{sec.con}

This paper adds to and extends on data of three important interstellar ice candidates:  acetaldehyde, ethanol, and dimethyl ether. A number of selected bands are fully characterized in FWHM and peak positions and show clear changes in various matrices. Our conclusions are summarized as follows:

\begin{enumerate}
\item The most promising bands to identify the COMs studied here in interstellar ice spectra are the 7.427 and 5.88 $\mu$m bands of acetaldehyde, the 7.240 and 11.36 $\mu$m bands of ethanol, and the 8.011 and 8.592 $\mu$m bands of dimethyl ether.
\item Matrix characteristic shifts in peak position and FWHM are seen for several bands. The acetaldehyde CH$_{3}$ deformation and CO stretching mode can be distinguished in the H$_{2}$O, CO, CH$_{3}$OH, and CO:CH$_{3}$OH matrices. Ethanol shows generally less distinctive shifts and only bands in the water matrix are unique. At low temperatures matrix specific dimethyl ether band shifts can be identified, specifically for the CH$_{3}$ rocking mode at 8.011 $\mu$m.
\item Given the higher complexity of the involved spectra, unambiguous identifications need to involve different bands that reflect bandwidths and intensity ratios as found in the laboratory studies. The dependence on matrix environment and temperature provides a tool to use these transitions as a remote diagnostic instrument.
\item Analysis of the ISO W33A spectrum in the 7 $\mu$m region shows a number of features that can be assigned to the COMs studied in this work. The 7.40 $\mu$m feature matches the position of the CH$_{3}$ s-deformation mode of acetaldehyde, and the 7.22 $\mu$m feature is plausibly caused by the CH$_3$ s-deformation mode of ethanol. It is likely that 7.22 $\mu$m band is specifically caused by ethanol mixed in water. Abundances of both molecules with respect to water ice are determined to be $\leq$2.3\% and $\leq$3.4\% for acetaldehyde and ethanol, respectively.
\end{enumerate}

\begin{acknowledgements}
The authors thank M.E. Palumbo for useful discussions on band profile changes due to grain shape differences. We also thank S. Ioppolo for many discussions. 
This research was funded through a VICI grant of NWO, the Netherlands Organization for Scientific Research; Astrochemistry in Leiden is supported by the European Union A-ERC grant 291141 CHEMPLAN, by the Netherlands Research School for Astronomy (NOVA), and by a Royal Netherlands Academy of Arts and Sciences (KNAW) professor prize. 
\end{acknowledgements}

%-------------------------------------------------------------------

\bibliographystyle{aa}
\bibliography{lib}

\appendix

%__________________________________________________________________

\clearpage

\section{Spectra}
\label{ap.spectra}

The following figures show the spectra of acetaldehyde, ethanol, and dimethylether mixed in water, CO, methanol, and CO:methanol in the range of 2.5 to 20.0 $\mu$m. All spectra are taken at 15~K. 

\begin{figure*}
\begin{center}
\includegraphics[width=\hsize]{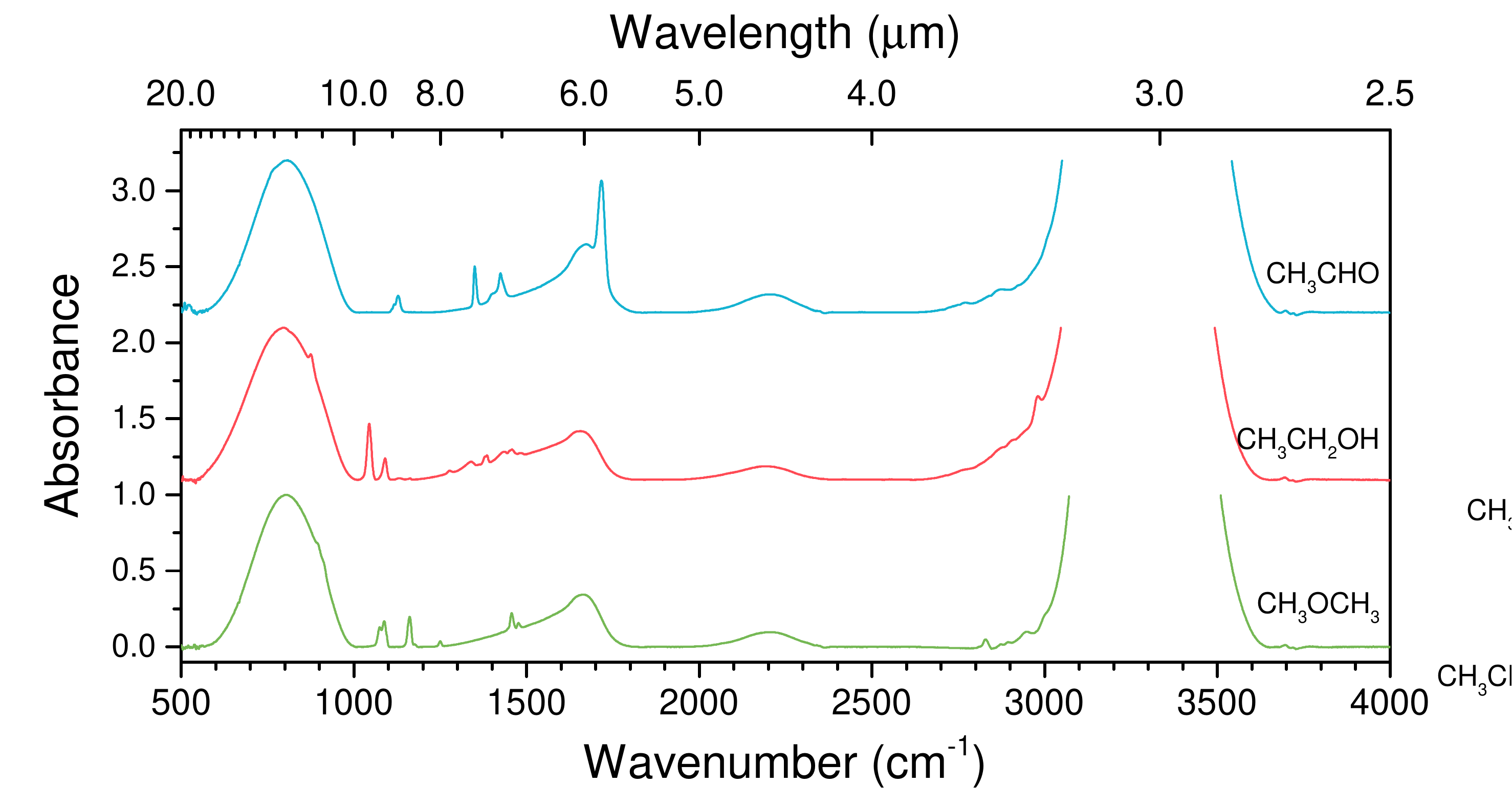}
\caption{Spectra of acetaldehyde (blue), ethanol (red), and dimethyl ether (green) mixed in water at 15~K in the range of 2.5 to 20.0 $\mu$m. }
\label{fig.COM_comp_15K_water}
\end{center}
\end{figure*}

\begin{figure*}
\begin{center}
\includegraphics[width=\hsize]{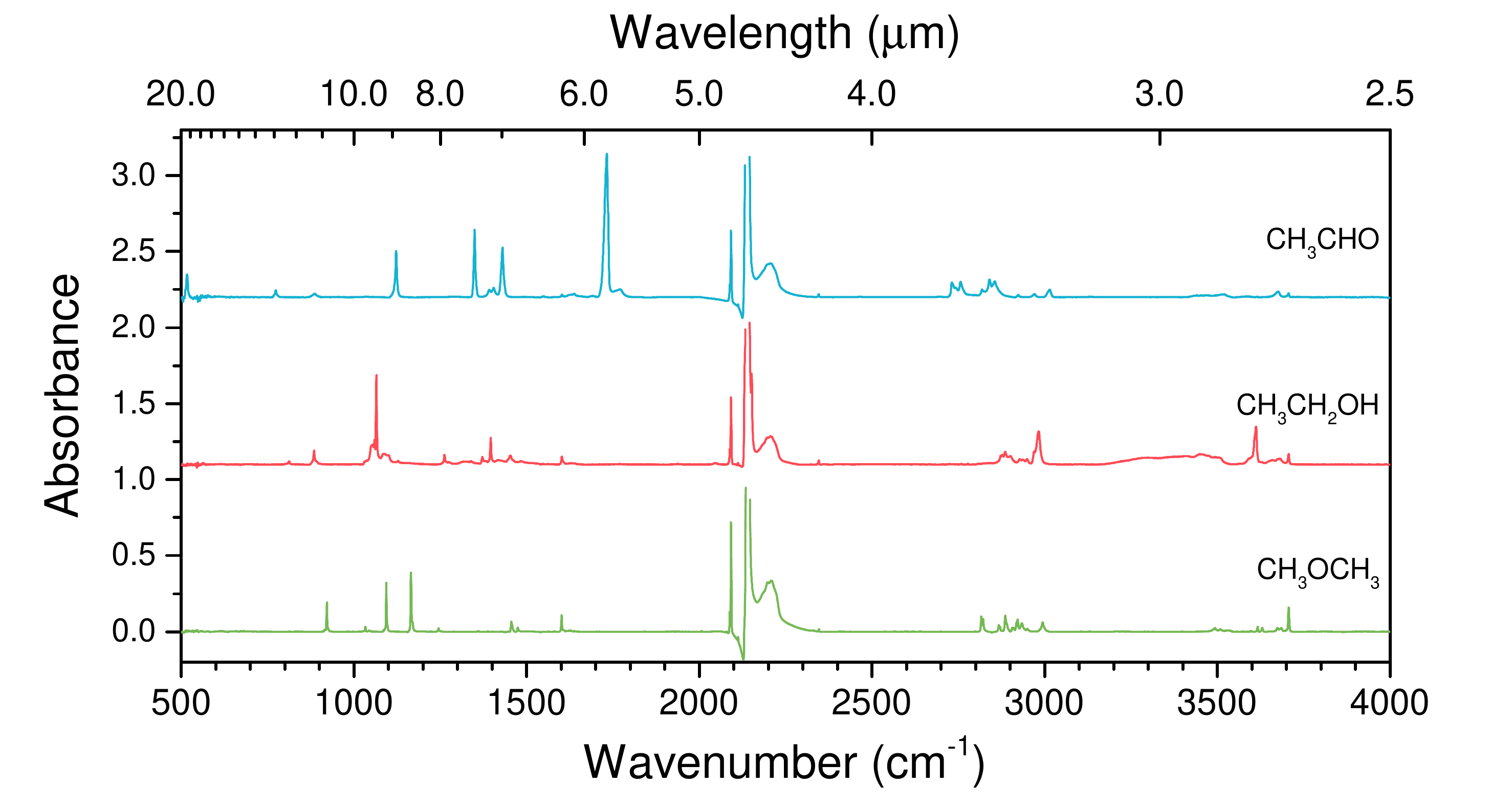}
\caption{Spectra of acetaldehyde (blue), ethanol (red), and dimethyl ether (green) mixed in CO at 15~K in the range of 2.5 to 20.0 $\mu$m. }
\label{fig.COM_comp_15K_CO}
\end{center}
\end{figure*}

\begin{figure*}
\begin{center}
\includegraphics[width=\hsize]{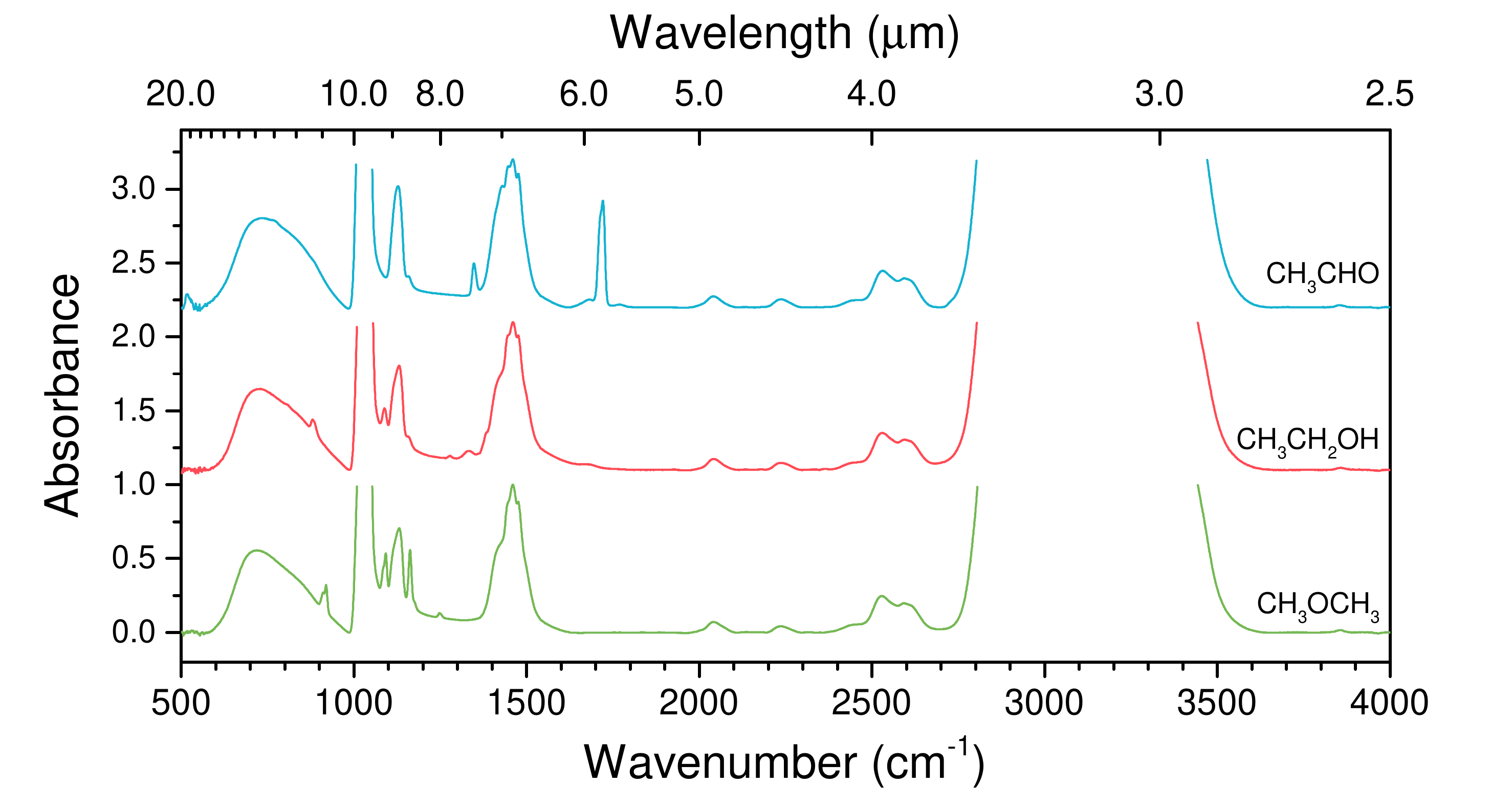}
\caption{Spectra of acetaldehyde (blue), ethanol (red), and dimethyl ether (green) mixed in methanol at 15~K in the range of 2.5 to 20.0 $\mu$m. }
\label{fig.COM_comp_15K_methanol}
\end{center}
\end{figure*}

\begin{figure*}
\begin{center}
\includegraphics[width=\hsize]{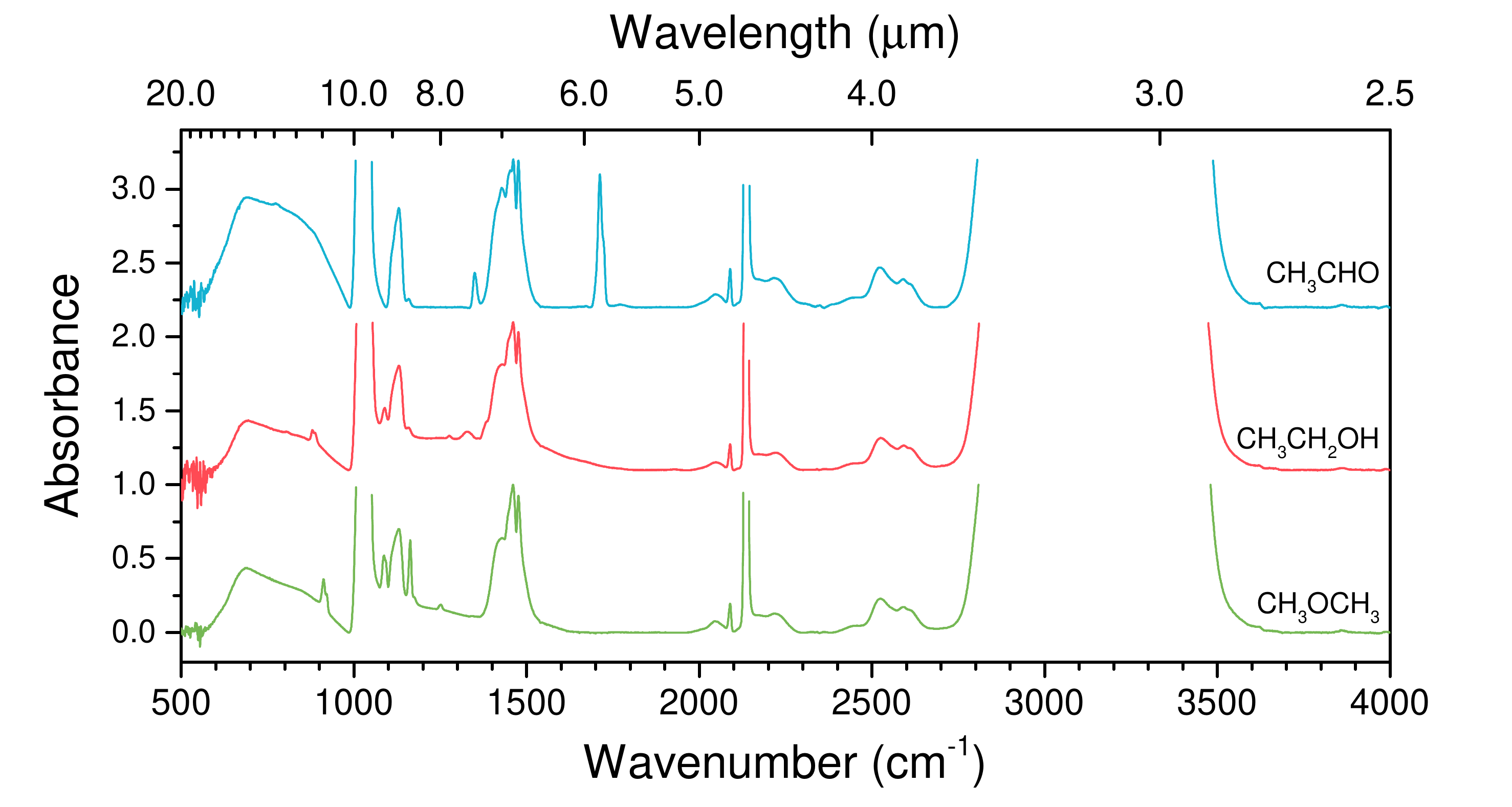}
\caption{Spectra of acetaldehyde (blue), ethanol (red), and dimethyl ether (green) mixed in CO:CH$_{3}$OH at 15~K in the range of 2.5 to 20.0 $\mu$m. }
\label{fig.COM_comp_15K_co_methanol}
\end{center}
\end{figure*}

\clearpage

\section{Overview of peak position, FWHM, and integrated absorbance ratios of selected transitions}
\label{ap.tables}

In this section tables are presented that list peak positions, FWHMs, and integrated absorbance ratios of selected acetaldehyde, ethanol, and dimethyl ether transitions. Where necessary, peak positions are given of both baseline corrected and matrix subtracted spectra. The peak position and FWHM are given in wavenumber (cm$^{-1}$) and wavelength ($\mu$m). Separate tables list the variation in band intensities over a range of temperatures for each mixture (e.g. Table \ref{tab.acetal_pure_area}). Values in these tables are usually normalized to the strongest transition at 15~K, which also remains identifiable over the entire temperature range. Exceptions are made for bands that are potentially in saturation, for example the CO stretching mode in pure acetaldehyde ice. 

In the tables various asterisks are used to indicate special circumstances. An asterick indicates that the FWHM is the result of two or more blended peaks. Double astericks indicate multiple peaks, which are often caused by a different matrix or surrounding interactions of the band. Occasionally the matrix cannot be properly subtracted from the feature under investigation, which results in FWHMs with higher uncertainty or in FWHMs that cannot be determined at all. Finally, a triple astericks indicates ice transitions that are thought to be strong enough to saturate the IR spectrometer signal. 

\subsection{Acetaldehyde}
\label{ap.acetaldehyde_spec}

\begin{table*}[ht]
\centering
\caption[]{Peak positions and FWHM of the acetaldehyde CH$_{3}$ rocking + CC stretching + CCO bending mode at 8.909 $\mu$m.}
\begin{tabular}{| l | c | l l | l l | l l |}
\hline
        \multirow{2}{*}{Mixture} & Temperature &  \multicolumn{2}{c |}{$\lambda_{\rm{peak,-baseline}}$} &  \multicolumn{2}{c |}{$\lambda_{\rm{peak,-matrix}}$} & \multicolumn{2}{c |}{FWHM} \\ 
         & (K) & (cm$^{-1}$) & ($\mu$m) & (cm$^{-1}$) & ($\mu$m) & (cm$^{-1}$) &  ($\mu$m) \\ 
        \hline
Pure    &       \multirow{6}{*}{15}     &       1122.4  &       8.9097  &       --      &       --      &       13.0    &       0.1032  \\
CH$_3$CHO :H$_2$O       &       &       1116.1  &       8.9598  &       1116.1  &       8.9598  &       --      &       --      \\      
        &       &       1127.7  &       8.8678  &       1127.7  &       8.8678  &       14.3    &       0.1121  \\      
CH$_3$CHO :CO   &       &       1121.9  &       8.9136  &       --      &       --      &       \phantom{1}5.0  &       0.0394  \\      
CH$_3$CHO :CH$_3$OH     &       &       --      &       --      &       --      &       --      &       --      &       --      \\      
CH$_3$CHO :CO:CH$_3$OH  &       &       --      &       --      &       --      &       --      &       --      &       --      \\
\end{tabular}
\tablefoot{ *FWHM result of two or more blended peaks. **FWHM uncertain/not determined owing to uncertain matrix subtraction. ***Transition likely saturated.}
\label{tab.acetal_cc_stretch}
\end{table*}

\begin{table*}
\centering
\caption[]{Peak position and FWHM of the acetaldehyde CH$_3$ s-deformation + CH waging mode at 7.427 $\mu$m.}
\begin{tabular}{| l | c | l l | l l | l l |}
\hline
        \multirow{2}{*}{Mixture} & Temperature &  \multicolumn{2}{c |}{$\lambda_{\rm{peak,-baseline}}$} &  \multicolumn{2}{c |}{$\lambda_{\rm{peak,-matrix}}$} & \multicolumn{2}{c |}{FWHM} \\ 
         & (K) & (cm$^{-1}$) & ($\mu$m) & (cm$^{-1}$) & ($\mu$m) & (cm$^{-1}$) &  ($\mu$m) \\ 
        \hline
CH$_3$CHO       &       \multirow{5}{*}{15}     &       1346.6  &       7.4264  &       --      &       --      &       13.5    &       0.0744* \\
CH$_3$CHO :H$_2$O       &       &       1349.9  &       7.4078  &       1349.9  &       7.4078  &       \phantom{1}9.2  &       0.0502  \\      
CH$_3$CHO :CO   &       &       1349.4  &       7.4104  &       --      &       --      &       \phantom{1}4.8  &       0.0262  \\      
CH$_3$CHO :CH$_3$OH     &       &       1347.5  &       7.4211  &       --      &       --      &       13.0    &       0.0714  \\      
CH$_3$CHO :CO:CH$_3$OH  &       &       1349.4  &       7.4105  &       --      &       --      &       12.6    &       0.0691  \\      
\hline
CH$_3$CHO       &       \multirow{5}{*}{30}     &       1346.1  &       7.4290  &       --      &       --      &       13.7    &       0.0754  \\
\end{tabular}
\tablefoot{ *FWHM result of two or more blended peaks. **FWHM uncertain/not determined owing to uncertain matrix subtraction. ***Transition likely saturated.}
\label{tab.acetal_ch3_sdeform}
\end{table*}

\begin{table*}
\centering
\caption[]{Peak position and FWHM of the acetaldehyde CH$_3$ deformation mode at 6.995 $\mu$m.}
\begin{tabular}{| l | c | l l | l l | l l |}
\hline
        \multirow{2}{*}{Mixture} & Temperature &  \multicolumn{2}{c |}{$\lambda_{\rm{peak,-baseline}}$} &  \multicolumn{2}{c |}{$\lambda_{\rm{peak,-matrix}}$} & \multicolumn{2}{c |}{FWHM} \\ 
         & (K) & (cm$^{-1}$) & ($\mu$m) & (cm$^{-1}$) & ($\mu$m) & (cm$^{-1}$) &  ($\mu$m) \\ 
        \hline
        CH$_3$CHO       &       \multirow{5}{*}{15}     &       1429.5  &       6.9956  &       --      &       --      &       22.6    &       0.1105  \\
CH$_3$CHO :H$_2$O       &       &       1424.2  &       7.0216  &       1424.2  &       7.0216  &       17.0    &       0.0837  \\      
CH$_3$CHO :CO   &       &       1430.4  &       6.9909  &       --      &       --      &       \phantom{1}7.7  &       0.0379  \\      
CH$_3$CHO :CH$_3$OH     &       &       --      &       --      &       --      &       --      &       --      &       --      \\      
CH$_3$CHO :CO:CH$_3$OH  &       &       --      &       --      &       --      &       --      &       --      &       --      \\      
\hline
CH$_3$CHO       &       \multirow{5}{*}{30}     &       1429.5  &       6.9956  &       --      &       --      &       22.8    &       0.1115  \\
\end{tabular}
\tablefoot{ *FWHM result of two or more blended peaks. **FWHM uncertain/not determined owing to uncertain matrix subtraction. ***Transition likely saturated.}
\label{tab.acetal_ch3_ddeform}
\end{table*}

\begin{table*}
\centering
\caption[]{Peak position and FWHM of the acetaldehyde CO stretching mode at 5.803 $\mu$m.}
\begin{tabular}{| l | c | l l | l l | l l |}
\hline
        \multirow{2}{*}{Mixture} & Temperature &  \multicolumn{2}{c |}{$\lambda_{\rm{peak,-baseline}}$} &  \multicolumn{2}{c |}{$\lambda_{\rm{peak,-matrix}}$} & \multicolumn{2}{c |}{FWHM} \\ 
         & (K) & (cm$^{-1}$) & ($\mu$m) & (cm$^{-1}$) & ($\mu$m) & (cm$^{-1}$) &  ($\mu$m) \\ 
        \hline
        CH$_3$CHO***    &       \multirow{5}{*}{15}     &       1723.6  &       5.8019  &       --      &       --      &       19.5    &       0.0659  \\
CH$_3$CHO :H$_2$O       &       &       1716.8  &       5.8247  &       1717.8  &       5.8215  &       20.9    &       0.0709  \\      
CH$_3$CHO :CO   &       &       1732.2  &       5.7729  &       --      &       --      &       12.6    &       0.0419  \\      
CH$_3$CHO :CH$_3$OH     &       &       1721.2  &       5.8101  &       --      &       --      &       20.4    &       0.0690  \\      
CH$_3$CHO :CO:CH$_3$OH  &       &       1712.5  &       5.8395  &       --      &       --      &       17.3    &       0.0589* \\      
\hline
CH$_3$CHO***    &       \multirow{6}{*}{30}     &       1723.6  &       5.8019  &       --      &       --      &       20.2    &       0.0682  \\
\end{tabular}
\tablefoot{ *FWHM result of two or more blended peaks. **FWHM uncertain/not determined owing to uncertain matrix subtraction. ***Transition likely saturated.}
\label{tab.acetal_co_stretch}
\end{table*}

\clearpage

\subsection{Acetaldehyde band areas}
\label{ap.acetaldehyde_area}

\begin{table*}
\centering
\caption{Integrated absorbance ratios of selected transitions in pure acetaldehyde.}
\begin{tabular}{| l | l | l | l | l |}
\hline
        Temperature & CH$_{3}$ rock. + CC stretch. & CH$_{3}$ deform. + CH wag. & CH$_{3}$ deform. & CO stretch. \\
         & + CCO bend. &  & &  \\
        (K) & 8.909$\mu$m & 7.427$\mu$m & 6.995$\mu$m & 5.803$\mu$m \\
        \hline
15      &       0.72    &       1.00    &       1.07    &       4.32    \\
30      &       0.73    &       1.00    &       1.07    &       4.36    \\
70      &       0.70    &       0.95    &       0.99    &       4.27    \\
90      &       0.59    &       0.85    &       0.87    &       4.16    \\
110     &       0.58    &       0.83    &       0.85    &       4.09    \\
120     &       0.57    &       0.81    &       0.82    &       4.01    \\
        \hline
        \noalign{\smallskip}
\end{tabular}
\tablefoot{ Owing to possible saturation of the C=O stretch mode the band intensities are normalized on the CH$_{3}$ s-deformation band at 15~K}
\label{tab.acetal_pure_area}
\end{table*}

\begin{table*}
\centering
\caption[]{Integrated absorbance ratios of selected transitions in acetaldehyde:H$_2$O.}
\begin{tabular}{| l | l | l | l | l |}
\hline
        Temperature & CH$_{3}$ rock. + CC stretch. & CH$_{3}$ deform. + CH wag. & CH$_{3}$ deform. & CO stretch. \\
         & + CCO bend. &  & &  \\
        (K) & 8.909$\mu$m & 7.427$\mu$m & 6.995$\mu$m & 5.803$\mu$m \\
        \hline
15      &       0.15    &       0.21    &       0.27    &       1.00    \\
30      &       0.16    &       0.21    &       0.25    &       1.02    \\
70      &       0.15    &       0.21    &       0.22    &       1.00    \\
90      &       0.15    &       0.03    &       0.22    &       1.01    \\
110     &       0.15    &       0.19    &       0.20    &       0.94    \\
120     &       0.15    &       0.18    &       0.19    &       0.83    \\
140     &       0.12    &       0.15    &       0.17    &       0.61    \\
160     &       0.11    &       0.13    &       0.12    &       0.57    \\
        \hline
        \noalign{\smallskip}
\end{tabular}

%\tablefoot{}
\label{tab.acetal_H2O_area}
\end{table*}

\begin{table*}
\centering
\caption[]{Integrated absorbance ratios of selected transitions in acetaldehyde:CO.}
\begin{tabular}{| l | l | l | l | l |}
\hline
        Temperature & CH$_{3}$ rock. + CC stretch. & CH$_{3}$ deform. + CH wag. & CH$_{3}$ deform. & CO stretch. \\
         & + CCO bend. &  & &  \\
        (K) & 8.909$\mu$m & 7.427$\mu$m & 6.995$\mu$m & 5.803$\mu$m \\
        \hline
15      &       0.18    &       0.21    &       0.25    &       1.00    \\
30      &       0.18    &       0.22    &       0.24    &       1.00    \\
        \hline
        \noalign{\smallskip}
\end{tabular}
%\tablefoot{}
\label{tab.acetal_co_area}
\end{table*}

\begin{table*}
\centering
\caption[]{Integrated absorbance ratios of selected transitions in acetaldehyde:CH$_{3}$OH.}
\begin{tabular}{| l | l | l | l | l |}
\hline
        Temperature & CH$_{3}$ rock. + CC stretch. & CH$_{3}$ deform. + CH wag. & CH$_{3}$ deform. & CO stretch. \\
         & + CCO bend. &  & &  \\
        (K) & 8.909$\mu$m & 7.427$\mu$m & 6.995$\mu$m & 5.803$\mu$m \\
        \hline
15      &       --      &       0.19    &       --      &       1.00    \\
30      &       --      &       0.19    &       --      &       0.99    \\
70      &       --      &       0.18    &       --      &       0.95    \\
90      &       --      &       0.17    &       --      &       0.94    \\
110     &       --      &       0.16    &       --      &       0.96    \\
120     &       --      &       --      &       --      &       0.62    \\
140     &       --      &       --      &       --      &       0.35    \\
        \hline
        \noalign{\smallskip}
\end{tabular}

%\tablefoot{}
\label{tab.acetal_CH$_{3}$OH_area}
\end{table*}

\begin{table*}
\centering
\caption[]{Integrated absorbance ratios of selected transitions in acetaldehyde:CO:CH$_{3}$OH.}
\begin{tabular}{| l | l | l | l | l |}
\hline
        Temperature & CH$_{3}$ rock. + CC stretch. & CH$_{3}$ deform. + CH wag. & CH$_{3}$ deform. & CO stretch. \\
         & + CCO bend. &  & &  \\
        (K) & 8.909$\mu$m & 7.427$\mu$m & 6.995$\mu$m & 5.803$\mu$m \\
        \hline
15      &       --      &       0.17    &       --      &       1.00    \\
30      &       --      &       0.17    &       --      &       0.94    \\
70      &       --      &       0.15    &       --      &       0.76    \\
90      &       --      &       0.14    &       --      &       0.78    \\
110     &       --      &       --      &       --      &       0.82    \\
120     &       --      &       --      &       --      &       0.48    \\
        \hline
        \noalign{\smallskip}
\end{tabular}

%\tablefoot{}
\label{tab.acetal_coCH$_{3}$OH_area}
\end{table*}

\clearpage

\subsection{Ethanol}
\label{ap.ethanol_spec}

\begin{table*}
\centering
\caption[]{Peak position and FWHM of the ethanol CC stretching mode at 11.36$\mu$m.}
\begin{tabular}{| l | c | l l | l l | l l |}
\hline
        \multirow{2}{*}{Mixture} & Temperature &  \multicolumn{2}{c |}{$\lambda_{\rm{peak,-baseline}}$} &  \multicolumn{2}{c |}{$\lambda_{\rm{peak,-matrix}}$} & \multicolumn{2}{c |}{FWHM} \\ 
         & (K) & (cm$^{-1}$) & ($\mu$m) & (cm$^{-1}$) & ($\mu$m) & (cm$^{-1}$) &  ($\mu$m) \\ 
        \hline
CH$_3$CH$_2$OH  &       \multirow{6}{*}{15}     &       879.9   &       11.3654 &       --      &       --      &       18.3    &       0.2344* \\
CH$_3$CH$_2$OH :H$_2$O  &       &       875.0   &       11.4280 &       877.0   &       11.4029 &       11.7    &       0.1526  \\      
CH$_3$CH$_2$OH :CO      &       &       884.7   &       11.3035 &       --      &       --      &       \phantom{1}3.8  &       0.0482  \\      
CH$_3$CH$_2$OH :CH$_3$OH        &       &       880.3   &       11.3592 &       882.3   &       11.3344 &       16.4    &       0.2105  \\      
CH$_3$CH$_2$OH :CO:CH$_3$OH     &       &       880.3   &       11.3592 &       880.8   &       11.3530 &       16.4    &       0.2097* \\      
        &       &       887.6   &       11.2666 &       888.1   &       11.2605 &       --      &       --      \\
\end{tabular}
\tablefoot{ *FWHM result of two or more blended peaks. **FWHM uncertain/not determined owing to uncertain matrix subtraction. ***Transition likely saturated.}
\label{tab.etoh_cc_stretch}
\end{table*}

\begin{table*}
\centering
\caption[]{Peak position and FWHM of the ethanol CO stretching mode at 9.514$\mu$m.}
\begin{tabular}{| l | c | l l | l l | l l |}
\hline
        \multirow{2}{*}{Mixture} & Temperature &  \multicolumn{2}{c |}{$\lambda_{\rm{peak,-baseline}}$} &  \multicolumn{2}{c |}{$\lambda_{\rm{peak,-matrix}}$} & \multicolumn{2}{c |}{FWHM} \\ 
         & (K) & (cm$^{-1}$) & ($\mu$m) & (cm$^{-1}$) & ($\mu$m) & (cm$^{-1}$) &  ($\mu$m) \\ 
        \hline
CH$_3$CH$_2$OH  &       \multirow{5}{*}{15}     &       1051.0  &       9.5146  &       --      &       --      &       19.0    &       0.1724  \\
CH$_3$CH$_2$OH :H$_2$O  &       &       1044.3  &       9.5761  &       1044.3  &       9.5761  &       14.3    &       0.1315  \\      
CH$_3$CH$_2$OH :CO      &       &       1065.0  &       9.3897  &       --      &       --      &       \phantom{1}2.9  &       0.0253  \\      
CH$_3$CH$_2$OH :CH$_3$OH        &       &       --      &       --      &       --      &       --      &       --      &       --      \\      
CH$_3$CH$_2$OH :CO:CH$_3$OH     &       &       --      &       --      &       --      &       --      &       --      &       --      \\      
\hline
CH$_3$CH$_2$OH  &       \multirow{5}{*}{30}     &       1051.0  &       9.5146  &       --      &       --      &       17.6    &       0.1598  \\
\end{tabular}
\tablefoot{ *FWHM result of two or more blended peaks. **FWHM uncertain/not determined owing to uncertain matrix subtraction. ***Transition likely saturated.}
\label{tab.etoh_co_stretch_1051}
\end{table*}

\begin{table*}
\centering
\caption[]{Peak position and FWHM of the ethanol CH$_{3}$ rocking mode at 9.170$\mu$m.}
\begin{tabular}{| l | c | l l | l l | l l |}
\hline
        \multirow{2}{*}{Mixture} & Temperature &  \multicolumn{2}{c |}{$\lambda_{\rm{peak,-baseline}}$} &  \multicolumn{2}{c |}{$\lambda_{\rm{peak,-matrix}}$} & \multicolumn{2}{c |}{FWHM} \\ 
         & (K) & (cm$^{-1}$) & ($\mu$m) & (cm$^{-1}$) & ($\mu$m) & (cm$^{-1}$) &  ($\mu$m) \\ 
        \hline
CH$_3$CH$_2$OH  &       \multirow{5}{*}{15}     &       1090.5  &       9.1697  &       --      &       --      &       24.4    &       0.2057  \\
CH$_3$CH$_2$OH :H$_2$O  &       &       1090.1  &       9.1738  &       1090.1  &       9.1738  &       12.9    &       0.1085  \\      
CH$_3$CH$_2$OH :CO      &       &       1084.8  &       9.2186  &       --      &       --      &       --      &       --      \\      
CH$_3$CH$_2$OH :CH$_3$OH        &       &       1088.6  &       9.1859  &       1088.6  &       9.1859  &       11.8    &       0.0996  \\      
CH$_3$CH$_2$OH :CO:CH$_3$OH     &       &       1089.6  &       9.1778  &       1087.7  &       9.1941  &       11.8    &       0.1001  \\      
\hline
CH$_3$CH$_2$OH  &       \multirow{5}{*}{30}     &       1091.0  &       9.1657  &       --      &       --      &       23.5    &       0.1978  \\
\end{tabular}
\tablefoot{ *FWHM result of two or more blended peaks. **FWHM uncertain/not determined owing to uncertain matrix subtraction. ***Transition likely saturated.}
\label{tab.etoh_co_stretch_1091}
\end{table*}

\begin{table*}
\centering
\caption[]{Peak position and FWHM of the ethanol CH$_{2}$ torsion mode at 7.842 $\mu$m.}
\begin{tabular}{| l | c | l l | l l | l l |}
\hline
        \multirow{2}{*}{Mixture} & Temperature &  \multicolumn{2}{c |}{$\lambda_{\rm{peak,-baseline}}$} &  \multicolumn{2}{c |}{$\lambda_{\rm{peak,-matrix}}$} & \multicolumn{2}{c |}{FWHM} \\ 
         & (K) & (cm$^{-1}$) & ($\mu$m) & (cm$^{-1}$) & ($\mu$m) & (cm$^{-1}$) &  ($\mu$m) \\ 
        \hline
CH$_3$CH$_2$OH  &       \multirow{6}{*}{15}     &       1275.2  &       7.8419  &       --      &       --      &       18.5    &       0.1133  \\
CH$_3$CH$_2$OH :H$_2$O  &       &       1278.1  &       7.8242  &       1276.2  &       7.8360  &       11.9    &       0.0728  \\      
CH$_3$CH$_2$OH :CO      &       &       1262.2  &       7.9228  &       --      &       --      &       \phantom{1}3.7  &       0.0231  \\      
        &       &       1273.8  &       7.8508  &       --      &       --      &       --      &       --      \\      
CH$_3$CH$_2$OH :CH$_3$OH        &       &       1278.6  &       7.8212  &       1278.6  &       7.8212  &       12.2    &       0.0746  \\      
CH$_3$CH$_2$OH :CO:CH$_3$OH     &       &       1276.2  &       7.8360  &       1276.2  &       7.8360  &       11.5    &       0.0707  \\
\end{tabular}
\tablefoot{ *FWHM result of two or more blended peaks. **FWHM uncertain/not determined owing to uncertain matrix subtraction. ***Transition likely saturated.}
\label{tab.etoh_mode1}
\end{table*}

\begin{table*}
\centering
\caption[]{Peak position and FWHM of the ethanol OH deformation mode at 7.518$\mu$m.}
\begin{tabular}{| l | c | l l | l l | l l |}
\hline
        \multirow{2}{*}{Mixture} & Temperature &  \multicolumn{2}{c |}{$\lambda_{\rm{peak,-baseline}}$} &  \multicolumn{2}{c |}{$\lambda_{\rm{peak,-matrix}}$} & \multicolumn{2}{c |}{FWHM} \\ 
         & (K) & (cm$^{-1}$) & ($\mu$m) & (cm$^{-1}$) & ($\mu$m) & (cm$^{-1}$) &  ($\mu$m) \\ 
        \hline
CH$_3$CH$_2$OH  &       \multirow{5}{*}{15}     &       1330.2  &       7.5179  &       --      &       --      &       50.9    &       0.2865  \\
CH$_3$CH$_2$OH :H$_2$O  &       &       1337.9  &       7.4745  &       1337.4  &       7.4772  &       30.1    &       0.1690  \\      
CH$_3$CH$_2$OH :CO      &       &       1339.3  &       7.4665  &       --      &       --      &       40.6    &       0.2306* \\      
CH$_3$CH$_2$OH :CH$_3$OH        &       &       1332.6  &       7.5043  &       1332.6  &       7.5043  &       33.6    &       0.1894  \\      
CH$_3$CH$_2$OH :CO:CH$_3$OH     &       &       1328.7  &       7.5261  &       1329.2  &       7.5233  &       33.7    &       0.1904  \\      
        \hline
CH$_3$CH$_2$OH  &       \multirow{5}{*}{30}     &       1330.2  &       7.5179  &       --      &       --      &       48.6    &       0.2726  \\
\end{tabular}
\tablefoot{ *FWHM result of two or more blended peaks. **FWHM uncertain/not determined owing to uncertain matrix subtraction. ***Transition likely saturated.}
\label{tab.etoh_mode2}
\end{table*}

\begin{table*}
\centering
\caption[]{Peak position and FWHM of the ethanol CH$_{3}$ s-deformation mode at 7.240$\mu$m.}
\begin{tabular}{| l | c | l l | l l | l l |}
\hline
        \multirow{2}{*}{Mixture} & Temperature &  \multicolumn{2}{c |}{$\lambda_{\rm{peak,-baseline}}$} &  \multicolumn{2}{c |}{$\lambda_{\rm{peak,-matrix}}$} & \multicolumn{2}{c |}{FWHM} \\ 
         & (K) & (cm$^{-1}$) & ($\mu$m) & (cm$^{-1}$) & ($\mu$m) & (cm$^{-1}$) &  ($\mu$m) \\ 
        \hline
CH$_3$CH$_2$OH  &       \multirow{7}{*}{15}     &       1381.3  &       7.2397  &       --      &       --      &       20.7    &       0.1088* \\
CH$_3$CH$_2$OH :H$_2$O  &       &       1379.8  &       7.2473  &       1379.3  &       7.2499  &       --      &       --      \\      
        &       &       1385.6  &       7.2171  &       1385.6  &       7.2171  &       15.6    &       0.0817  \\      
CH$_3$CH$_2$OH :CO      &       &       1372.1  &       7.2881  &       --      &       --      &       \phantom{1}5.7  &       0.0304  \\      
        &       &       1396.2  &       7.1622  &       --      &       --      &       \phantom{1}3.4  &       0.0175  \\      
CH$_3$CH$_2$OH :CH$_3$OH        &       &       1383.7  &       7.2271  &       1381.3  &       7.2397  &       14.9    &       0.0782**        \\
\end{tabular}
\tablefoot{ *FWHM result of two or more blended peaks. **FWHM uncertain/not determined owing to uncertain matrix subtraction. ***Transition likely saturated.}
\label{tab.etoh_mode3}
\end{table*}

\clearpage

\subsection{Ethanol normalized band areas}
\label{ap.ethanol_area}

\begin{table*}
\centering
\caption[]{Integrated absorbance ratios of selected transitions in pure ethanol.}
\begin{tabular}{| l | l | l | l | l | l | l |}
\hline
        Temperature & CC stretch. & CO stretch. & CH$_{3}$ rock. & CH$_{2}$ tors. & OH deform. & CH$_{3}$ s-deform. \\
        (K) & 11.36$\mu$m & 9.514$\mu$m & 9.170$\mu$m & 7.842$\mu$m & 7.518$\mu$m & 7.240$\mu$m \\
        \hline
15      &       0.23    &       1.00    &       0.59    &       0.06    &       0.35    &       0.20    \\
30      &       0.24    &       1.00    &       0.63    &       0.06    &       0.34    &       0.20    \\
70      &       0.24    &       1.00    &       0.65    &       0.06    &       0.38    &       0.23    \\
100     &       0.25    &       1.03    &       0.66    &       0.06    &       0.42    &       0.25    \\
120     &       0.28    &       1.04    &       0.64    &       0.05    &       0.47    &       0.23    \\
130     &       0.29    &       1.03    &       0.64    &       0.04    &       0.43    &       0.20    \\
140     &       0.28    &       1.02    &       0.64    &       0.04    &       0.44    &       0.21    \\
150     &       0.26    &       0.96    &       0.59    &       0.04    &       0.41    &       0.20    \\
        \hline
        \noalign{\smallskip}
\end{tabular}

\emph{\rm }
\label{tab.etoh_pure_area}
\end{table*}

\begin{table*}
\centering
\caption[]{Integrated absorbance ratios of selected transitions in ethanol:H$_2$O.}
\begin{tabular}{| l | l | l | l | l | l | l |}
\hline
        Temperature & CC stretch. & CO stretch. & CH$_{3}$ rock. & CH$_{2}$ tors. & OH deform. & CH$_{3}$ s-deform. \\
        (K) & 11.36$\mu$m & 9.514$\mu$m & 9.170$\mu$m & 7.842$\mu$m & 7.518$\mu$m & 7.240$\mu$m \\
        \hline
15      &       0.16    &       1.00    &       0.35    &       0.04    &       0.24    &       0.17    \\
30      &       0.17    &       1.01    &       0.35    &       0.05    &       0.28    &       0.19    \\
70      &       0.16    &       1.01    &       0.34    &       0.05    &       0.28    &       0.20    \\
100     &       0.12    &       0.97    &       0.33    &       0.05    &       0.30    &       0.21    \\
120     &       0.12    &       0.93    &       0.33    &       0.05    &       0.30    &       0.21    \\
130     &       0.12    &       0.93    &       0.33    &       0.04    &       0.27    &       0.21    \\
140     &       0.11    &       0.88    &       0.35    &       0.04    &       0.26    &       0.20    \\
150     &       0.11    &       0.86    &       0.36    &       0.04    &       0.21    &       0.16    \\
160     &       0.09    &       0.74    &       0.32    &       0.03    &       0.17    &       0.14    \\
        \hline
        \noalign{\smallskip}
\end{tabular}

\emph{\rm }
\label{tab.etoh_h2o_area}
\end{table*}

\begin{table*}
\centering
\caption[]{Integrated absorbance ratios of selected transitions in ethanol:CH$_{3}$OH.}
\begin{tabular}{| l | l | l | l | l | l | l |}
\hline
        Temperature & CC stretch. & CO stretch. & CH$_{3}$ rock. & CH$_{2}$ tors. & OH deform. & CH$_{3}$ s-deform. \\
        (K) & 11.36$\mu$m & 9.514$\mu$m & 9.170$\mu$m & 7.842$\mu$m & 7.518$\mu$m & 7.240$\mu$m \\
        \hline
15      &       1.00    &       --      &       0.99    &       0.16    &       1.50    &       0.57    \\
30      &       1.00    &       --      &       1.09    &       0.16    &       1.83    &       0.62    \\
70      &       1.01    &       --      &       1.38    &       0.19    &       3.14    &       0.81    \\
100     &       1.00    &       --      &       1.36    &       0.18    &       2.38    &       0.66    \\
120     &       1.05    &       --      &       1.12    &       0.18    &       --              &       0.85    \\
130     &       0.97    &       --      &       0.92    &       0.10    &       --              &       0.67    \\
140     &       0.98    &       --      &       1.12    &       0.09    &       --              &       0.66    \\
150     &       0.90    &       --      &       0.99    &       0.10    &       --              &       0.54    \\
        \hline
        \noalign{\smallskip}
\end{tabular}

\emph{\rm }
\label{tab.etoh_CH$_{3}$OH_area}
\end{table*}

\begin{table*}
\centering
\caption[]{Integrated absorbance ratios of selected transitions in ethanol:CO:CH$_{3}$OH.}
\begin{tabular}{| l | l | l | l | l | l | l |}
\hline
        Temperature & CC stretch. & CO stretch. & CH$_{3}$ rock. & CH$_{2}$ tors. & OH deform. & CH$_{3}$ s-deform. \\
        (K) & 11.36$\mu$m & 9.514$\mu$m & 9.170$\mu$m & 7.842$\mu$m & 7.518$\mu$m & 7.240$\mu$m \\
        \hline
15      &       1.00    &       --      &       0.87    &       0.19    &       1.73    &       0.66    \\
30      &       1.00    &       --      &       1.09    &       0.19    &       1.89    &       0.91    \\
70      &       0.94    &       --      &       1.37    &       0.19    &       2.80    &       0.79    \\
100     &       0.94    &       --      &       1.39    &       0.23    &       2.40    &       0.69    \\
120     &       0.77    &       --      &       1.03    &       0.11    &       --              &       0.61    \\
130     &       0.87    &       --      &       1.20    &       0.11    &       --              &       0.62    \\
140     &       0.90    &       --      &       1.16    &       0.12    &       --              &       0.57    \\
150     &       0.89    &       --      &       1.06    &       0.12    &       --              &       0.54    \\
        \hline
        \noalign{\smallskip}
\end{tabular}

\emph{\rm }
\label{tab.etoh_comethanol_area}
\end{table*}

\clearpage

\subsection{Dimethyl ether}
\label{ap.dimethylether_spec}

\begin{table*}
\centering
\caption[]{Peak position and FWHM of the dimethyl ether COC stretching mode at 10.85$\mu$m.}
\begin{tabular}{| l | c | l l | l l | l l |}
\hline
        \multirow{2}{*}{Mixture} & Temperature &  \multicolumn{2}{c |}{$\lambda_{\rm{peak,-baseline}}$} &  \multicolumn{2}{c |}{$\lambda_{\rm{peak,-matrix}}$} & \multicolumn{2}{c |}{FWHM} \\ 
         & (K) & (cm$^{-1}$) & ($\mu$m) & (cm$^{-1}$) & ($\mu$m) & (cm$^{-1}$) &  ($\mu$m) \\ 
        \hline
CH$_3$OCH$_3$   &       \multirow{8}{*}{15}     &       921.8   &       10.8483 &       --      &       --      &       10.6    &       0.1245  \\
CH$_3$OCH$_3$ :H$_2$O   &       &       898.2   &       11.1336 &       898.2   &       11.1336 &       --      &       --      \\      
        &       &       913.1   &       10.9514 &       913.1   &       10.9514 &       27.9    &       0.3406* \\      
CH$_3$OCH$_3$ :CO       &       &       921.8   &       10.8483 &       --      &       --      &       \phantom{1}2.1  &       0.0248  \\      
CH$_3$OCH$_3$ :CH$_3$OH &       &       911.2   &       10.9745 &       911.7   &       10.9687 &       --      &       --      \\      
        &       &       919.4   &       10.8767 &       919.4   &       10.8767 &       15.5    &       0.1846* \\
\end{tabular}
\tablefoot{ *FWHM result of two or more blended peaks. **FWHM uncertain/not determined owing to uncertain matrix subtraction. ***Transition likely saturated.}
\label{tab.dme_co_sstretch}
\end{table*}

\begin{table*}
\centering
\caption[]{Peak position and FWHM of the dimethyl ether COC stretching + CH$_{3}$ rocking mode at 9.141$\mu$m.}
\begin{tabular}{| l | c | l l | l l | l l |}
\hline
        \multirow{2}{*}{Mixture} & Temperature &  \multicolumn{2}{c |}{$\lambda_{\rm{peak,-baseline}}$} &  \multicolumn{2}{c |}{$\lambda_{\rm{peak,-matrix}}$} & \multicolumn{2}{c |}{FWHM} \\ 
         & (K) & (cm$^{-1}$) & ($\mu$m) & (cm$^{-1}$) & ($\mu$m) & (cm$^{-1}$) &  ($\mu$m) \\ 
        \hline
CH$_3$OCH$_3$   &       \multirow{6}{*}{15}     &       1094.4  &       9.1374  &       --      &       --      &       10.9    &       0.0912  \\
CH$_3$OCH$_3$ :H$_2$O   &       &       1074.6  &       9.3055  &       1074.6  &       9.3055  &       --      &       --      \\      
        &       &       1087.7  &       9.1941  &       1088.1  &       9.1900  &       24.7    &       0.2111* \\      
CH$_3$OCH$_3$ :CO       &       &       1093.9  &       9.1414  &       --      &       --      &       \phantom{1}2.0  &       0.0168  \\      
CH$_3$OCH$_3$ :CH$_3$OH &       &       1092.0  &       9.1576  &       1092.0  &       9.1576  &       12.5    &       0.1055* \\      
CH$_3$OCH$_3$ :CO:CH$_3$OH      &       &       1087.2  &       9.1982  &       1087.2  &       9.1982  &       13.8    &       0.1162  \\
\end{tabular}
\tablefoot{ *FWHM result of two or more blended peaks. **FWHM uncertain/not determined   owing to uncertain matrix subtraction. ***Transition likely saturated.}
\label{tab.dme_co_astretch}
\end{table*}

\begin{table*}
\centering
\caption[]{Peak position and FWHM of the dimethyl ether COC stretching and CH$_3$ rocking mode at 8.592$\mu$m.}
\begin{tabular}{| l | c | l l | l l | l l |}
\hline
        \multirow{2}{*}{Mixture} & Temperature &  \multicolumn{2}{c |}{$\lambda_{\rm{peak,-baseline}}$} &  \multicolumn{2}{c |}{$\lambda_{\rm{peak,-matrix}}$} & \multicolumn{2}{c |}{FWHM} \\ 
         & (K) & (cm$^{-1}$) & ($\mu$m) & (cm$^{-1}$) & ($\mu$m) & (cm$^{-1}$) &  ($\mu$m) \\ 
        \hline
CH$_3$OCH$_3$   &       \multirow{5}{*}{15}     &       1164.3  &       8.5888  &       --      &       --      &       \phantom{1}9.7  &       0.0714  \\
CH$_3$OCH$_3$ :H$_2$O   &       &       1161.4  &       8.6102  &       1161.4  &       8.6102  &       12.5    &       0.0924  \\      
CH$_3$OCH$_3$ :CO       &       &       1165.3  &       8.5817  &       --      &       --      &       \phantom{1}2.4  &       0.0177  \\      
CH$_3$OCH$_3$ :CH$_3$OH &       &       1162.9  &       8.5994  &       1162.9  &       8.5994  &       \phantom{1}9.6  &       0.0712  \\      
CH$_3$OCH$_3$ :CO:CH$_3$OH      &       &       1163.3  &       8.5959  &       1163.3  &       8.5959  &       \phantom{1}8.2  &       0.0608  \\      
        \hline
CH$_3$OCH$_3$   &       \multirow{5}{*}{30}     &       1163.8  &       8.5923  &       --      &       --      &       \phantom{1}9.7  &       0.0712  \\
\end{tabular}
\tablefoot{ *FWHM result of two or more blended peaks. **FWHM uncertain/not determined owing to uncertain matrix subtraction. ***Transition likely saturated.}
\label{tab.dme_ch3_rock_1164}
\end{table*}

\begin{table*}
\centering
\caption[]{Peak position and FWHM of the dimethyl ether CH$_3$ rocking mode at 8.011$\mu$m.}
\begin{tabular}{| l | c | l l | l l | l l |}
\hline
        \multirow{2}{*}{Mixture} & Temperature &  \multicolumn{2}{c |}{$\lambda_{\rm{peak,-baseline}}$} &  \multicolumn{2}{c |}{$\lambda_{\rm{peak,-matrix}}$} & \multicolumn{2}{c |}{FWHM} \\ 
         & (K) & (cm$^{-1}$) & ($\mu$m) & (cm$^{-1}$) & ($\mu$m) & (cm$^{-1}$) &  ($\mu$m) \\ 
        \hline
CH$_3$OCH$_3$   &       \multirow{5}{*}{15}     &       1248.2  &       8.0115  &       --      &       --      &       \phantom{1}9.6  &       0.0617  \\
CH$_3$OCH$_3$ :H$_2$O   &       &       1250.1  &       7.9992  &       1250.1  &       7.9992  &       \phantom{1}8.4  &       0.0540  \\      
CH$_3$OCH$_3$ :CO       &       &       1244.8  &       8.0333  &       --      &       --      &       \phantom{1}2.8  &       0.0180  \\      
CH$_3$OCH$_3$ :CH$_3$OH &       &       1247.7  &       8.0146  &       1247.7  &       8.0146  &       11.9    &       0.0761  \\      
CH$_3$OCH$_3$ :CO:CH$_3$OH      &       &       1251.6  &       7.9899  &       1251.6  &       7.9899  &       11.4    &       0.0728  \\      
        \hline
CH$_3$OCH$_3$   &       \multirow{5}{*}{30}     &       1248.2  &       8.0115  &       --      &       --      &       \phantom{1}9.6  &       0.0614  \\
\end{tabular}
\tablefoot{ *FWHM result of two or more blended peaks. **FWHM uncertain/not determined owing to uncertain matrix subtraction. ***Transition likely saturated.}
\label{tab.dme_ch3_rock_1248}
\end{table*}

\clearpage

\subsection{Dimethyl ether band areas}
\label{ap.dimethylether_area}

\begin{table*}
\centering
\caption[]{Integrated absorbance ratios of selected transitions in pure dimethyl ether.}
\begin{tabular}{| l | l | l | l | l |}
\hline
        Temperature & COC stretch. & COC stretch. + CH$_{3}$ rock. & COC stretch. + CH$_{3}$ rock. & CH$_{3}$ rock. \\
        (K) & 10.85$\mu$m & 9.141$\mu$m & 8.592$\mu$m & 8.011$\mu$m \\
        \hline
15      &       0.49    &       0.83    &       1.00    &       0.06    \\
30      &       0.49    &       0.83    &       0.99    &       0.07    \\
70      &       0.48    &       0.79    &       0.85    &       0.06    \\
90      &       0.48    &       0.78    &       0.84    &       0.06    \\
100     &       0.46    &       0.74    &       0.80    &       0.06    \\
        \hline
        \noalign{\smallskip}
\end{tabular}

\emph{\rm }
\label{tab.dme_pure_area}
\end{table*}

\begin{table*}
\centering
\caption[]{Integrated absorbance ratios of selected transitions in dimethyl ether:H$_2$O.}
\begin{tabular}{| l | l | l | l | l |}
\hline
        Temperature & COC stretch. & COC stretch. + CH$_{3}$ rock. & COC stretch. + CH$_{3}$ rock. & CH$_{3}$ rock. \\
        (K) & 10.85$\mu$m & 9.141$\mu$m & 8.592$\mu$m & 8.011$\mu$m \\
        \hline
15      &       0.36    &       1.00    &       0.74    &       0.08    \\
30      &       0.40    &       0.99    &       0.80    &       0.08    \\
70      &       0.38    &       0.94    &       0.79    &       0.07    \\
90      &       0.37    &       0.91    &       0.78    &       0.07    \\
100     &       0.34    &       0.89    &       0.78    &       0.07    \\
120     &       0.30    &       0.84    &       0.80    &       0.06    \\
140     &       0.17    &       0.71    &       0.79    &       0.05    \\
160     &       0.10    &       0.61    &       0.73    &       0.04    \\

        \hline
        \noalign{\smallskip}
\end{tabular}

\emph{\rm }
\label{tab.dme_h2o_area}
\end{table*}

\begin{table*}
\centering
\caption[]{Integrated absorbance ratios of selected transitions in dimethyl ether:CO.}
\begin{tabular}{| l | l | l | l | l |}
\hline
        Temperature & COC stretch. & COC stretch. + CH$_{3}$ rock. & COC stretch. + CH$_{3}$ rock. & CH$_{3}$ rock. \\
        (K) & 10.85$\mu$m & 9.141$\mu$m & 8.592$\mu$m & 8.011$\mu$m \\
        \hline
15      &       0.45    &       0.66    &       1.00    &       0.06    \\
30      &       0.45    &       0.70    &       1.00    &       0.07    \\
        \hline
        \noalign{\smallskip}
\end{tabular}

\emph{\rm }
\label{tab.dme_co_area}
\end{table*}

\begin{table*}
\centering
\caption[]{Integrated absorbance ratios of selected transitions in dimethyl ether:CH$_{3}$OH.}
\begin{tabular}{| l | l | l | l | l |}
\hline
        Temperature & COC stretch. & COC stretch. + CH$_{3}$ rock. & COC stretch. + CH$_{3}$ rock. & CH$_{3}$ rock. \\
        (K) & 10.85$\mu$m & 9.141$\mu$m & 8.592$\mu$m & 8.011$\mu$m \\
        \hline
15      &       0.71    &       0.71    &       1.00    &       0.10    \\
30      &       0.69    &       0.77    &       1.04    &       0.09    \\
70      &       0.65    &       0.83    &       1.04    &       0.09    \\
90      &       0.63    &       0.84    &       1.08    &       0.08    \\
100     &       0.60    &       0.82    &       1.05    &       0.08    \\
120     &       0.29    &       0.39    &       0.38    &       0.04    \\
        \hline
        \noalign{\smallskip}
\end{tabular}

\emph{\rm }
\label{tab.dme_CH$_{3}$OH_area}
\end{table*}

\begin{table*}
\centering
\caption[]{Integrated absorbance ratios of selected transitions in dimethyl ether:CO:CH$_{3}$OH.}
\begin{tabular}{| l | l | l | l | l |}
\hline
        Temperature & COC stretch. & COC stretch. + CH$_{3}$ rock. & COC stretch. + CH$_{3}$ rock. & CH$_{3}$ rock. \\
        (K) & 10.85$\mu$m & 9.141$\mu$m & 8.592$\mu$m & 8.011$\mu$m \\
        \hline
15      &       0.81    &       0.88    &       1.00    &       0.13    \\
30      &       0.83    &       1.02    &       1.13    &       0.12    \\
70      &       0.63    &       0.86    &       1.04    &       0.08    \\
90      &       0.63    &       0.83    &       1.00    &       0.08    \\
100     &       0.51    &       0.67    &       0.77    &       0.07    \\
        \hline
        \noalign{\smallskip}
\end{tabular}

\emph{\rm }
\label{tab.dme_coCH$_{3}$OH_area}
\end{table*}

\clearpage

%__________________________________________________________________
\twocolumn
\section{Spectroscopic data of selected bands}
\label{ap.visual}

\subsection{Acetaldehyde}

\begin{figure*}
\includegraphics[height=\hsize]{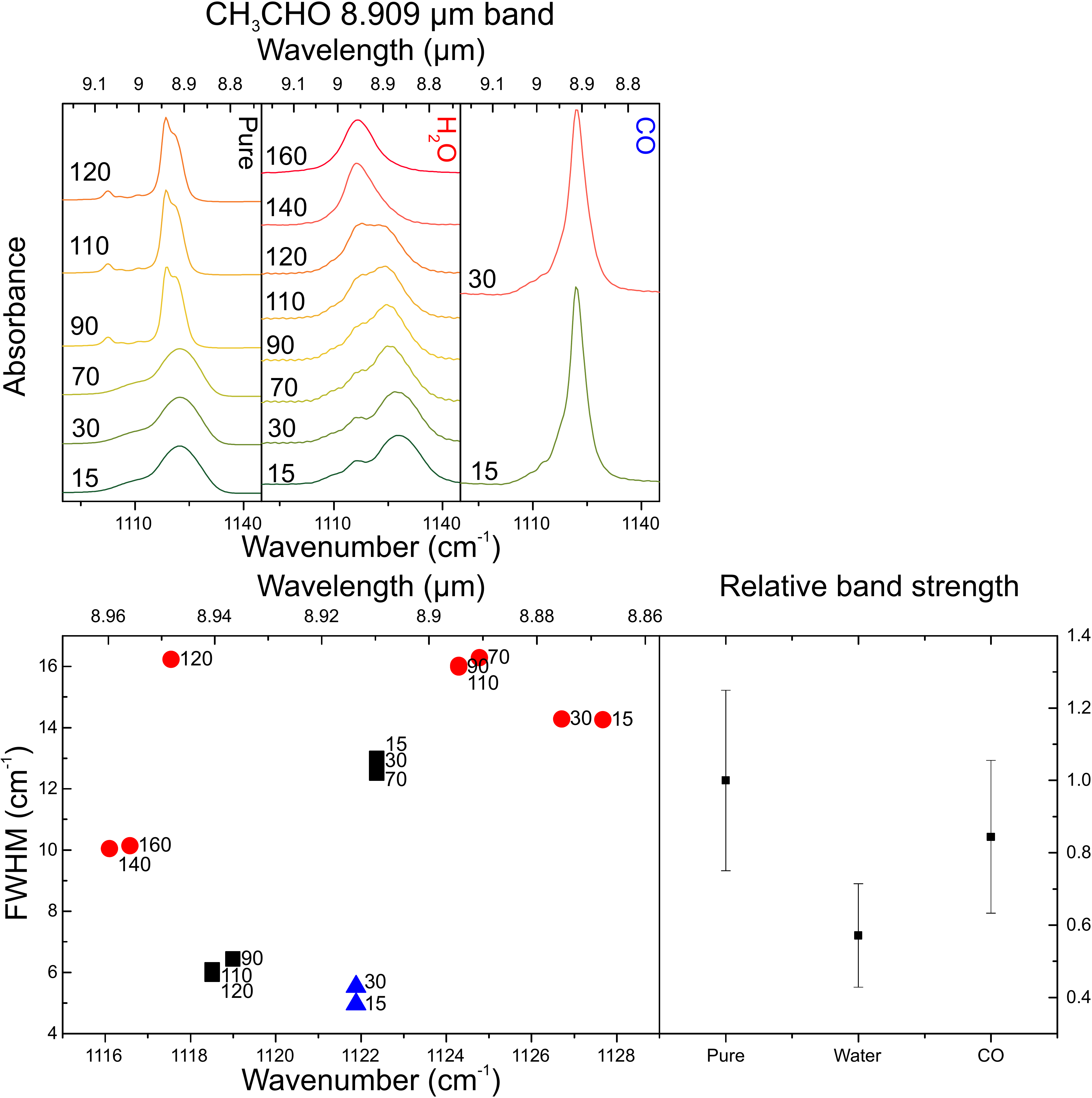}
\caption{Top: from left to right the acetaldehyde 8.909 $\mu$m band pure (black) and in water (red) and CO (blue) at various temperatures. Bottom left: peak position vs. FWHM plot, using the same colour coding. Bottom right: the relative band strength for the 8.909 $\mu$m band at 15~K in various matrices.}
\label{fig.A1120}
\end{figure*}

\begin{figure*}
\includegraphics[height=\hsize]{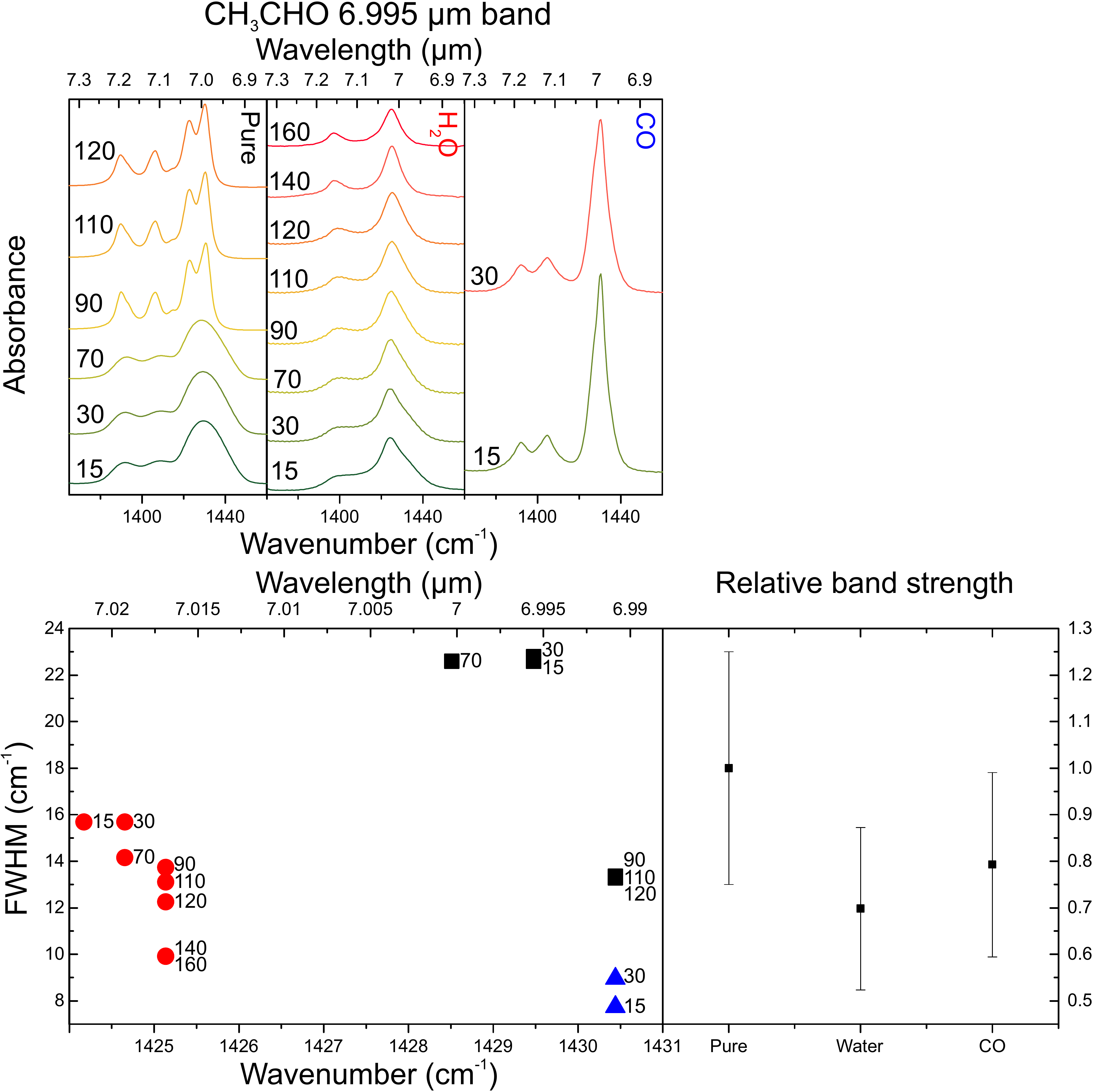}
\caption{Top: from left to right the acetaldehyde 6.995 $\mu$m band pure (black) and in water (red) and CO (blue) at various temperatures. Bottom left: peak position vs. FWHM plot, using the same colour coding. Bottom right: the relative band strength for the 6.995 $\mu$m band at 15~K in various matrices.}
\label{fig.A1430}
\end{figure*}

\begin{figure*}
\includegraphics[height=\hsize]{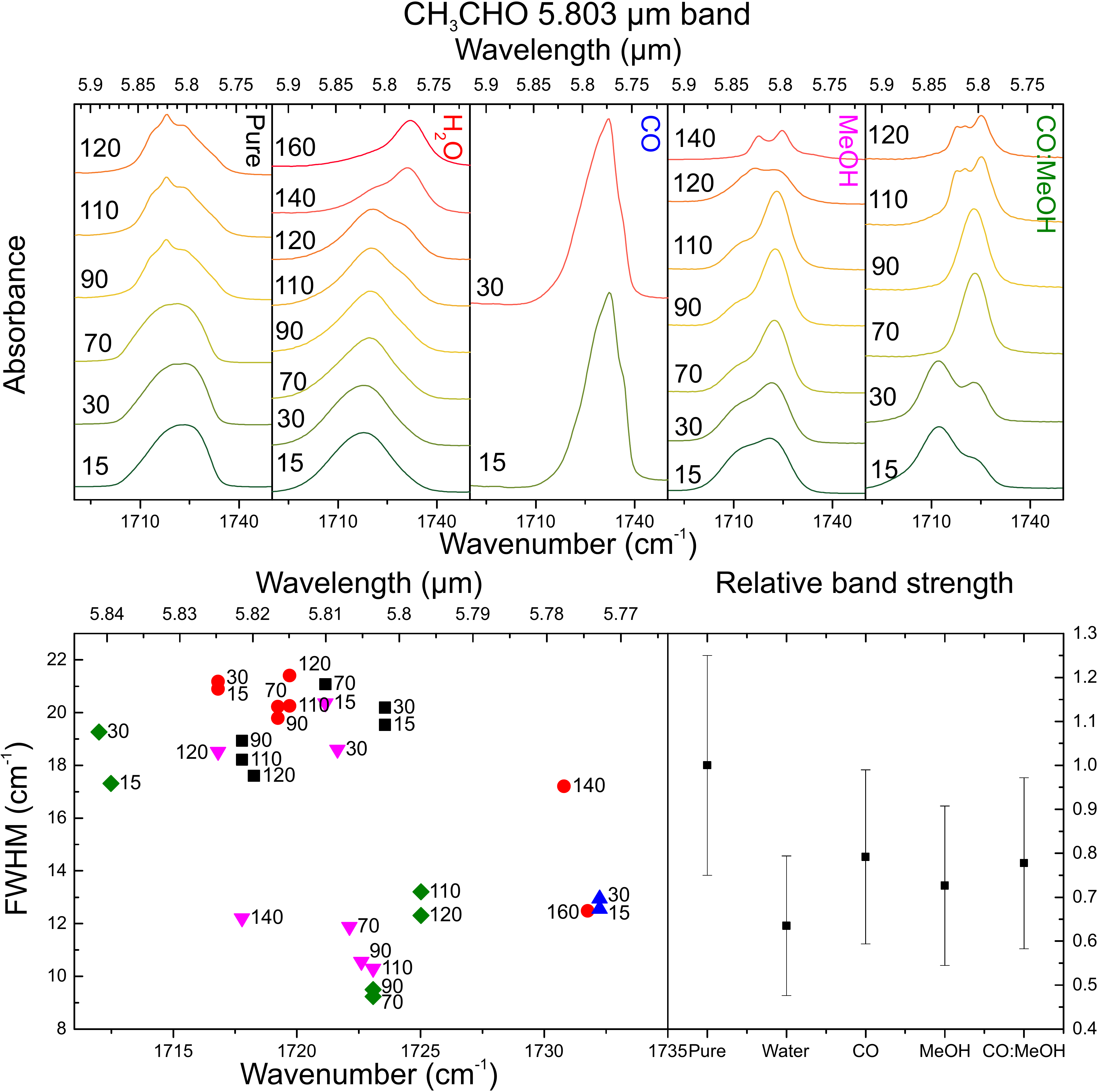}
\caption{Top: from left to right the acetaldehyde 5.803 $\mu$m band pure (black) and in water (red), CO (blue), methanol (purple), and CO:CH$_{3}$OH (green) at various temperatures. Bottom left: peak position vs. FWHM plot, using the same colour coding. Bottom right:\ the relative band strength for the 5.803 $\mu$m band at 15~K in various matrices.}
\label{fig.A1720}
\end{figure*}

\subsection{Ethanol}

\begin{figure*}
\includegraphics[height=\hsize]{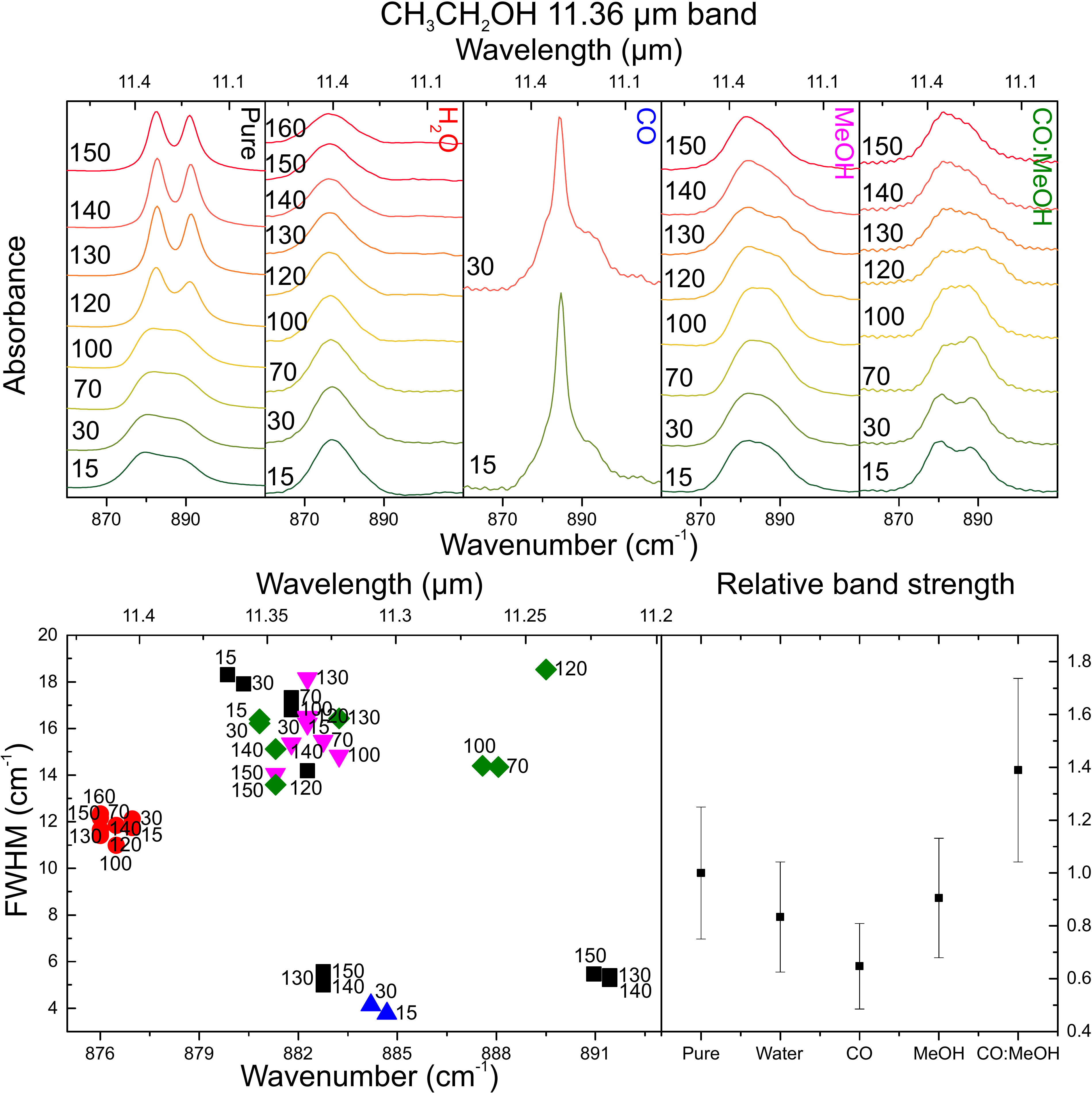}
\caption{Top: from left to right the ethanol 11.36 $\mu$m band pure (black) and in water (red), CO (blue), methanol (purple), and CO:CH$_{3}$OH (green) at various temperatures. Bottom left: peak position vs. FWHM plot, using the same colour coding. Bottom right: the relative band strength for the 11.36 $\mu$m band at 15~K in various matrices.}
\label{fig.E833}
\end{figure*}

\begin{figure*}
\includegraphics[height=\hsize]{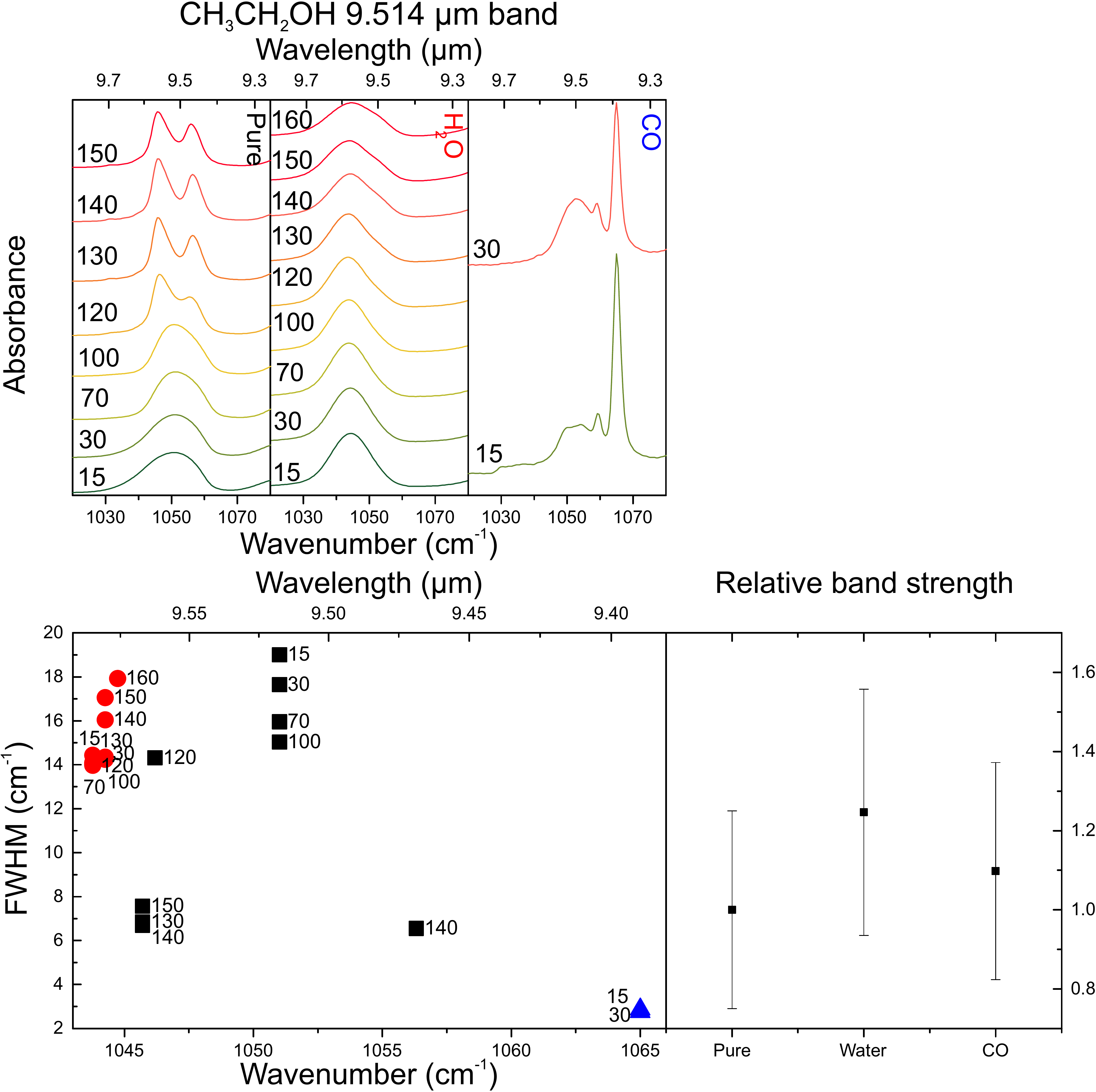}
\caption{Top: from left to right the ethanol 9.514 $\mu$m band pure (black) and in water (red) and CO (blue) at various temperatures. Bottom left: peak position vs. FWHM plot, using the same colour coding. Bottom right: the relative band strength for the 9.514 $\mu$m band at 15~K in various matrices.}
\label{fig.E1046}
\end{figure*}

\begin{figure*}
\includegraphics[height=\hsize]{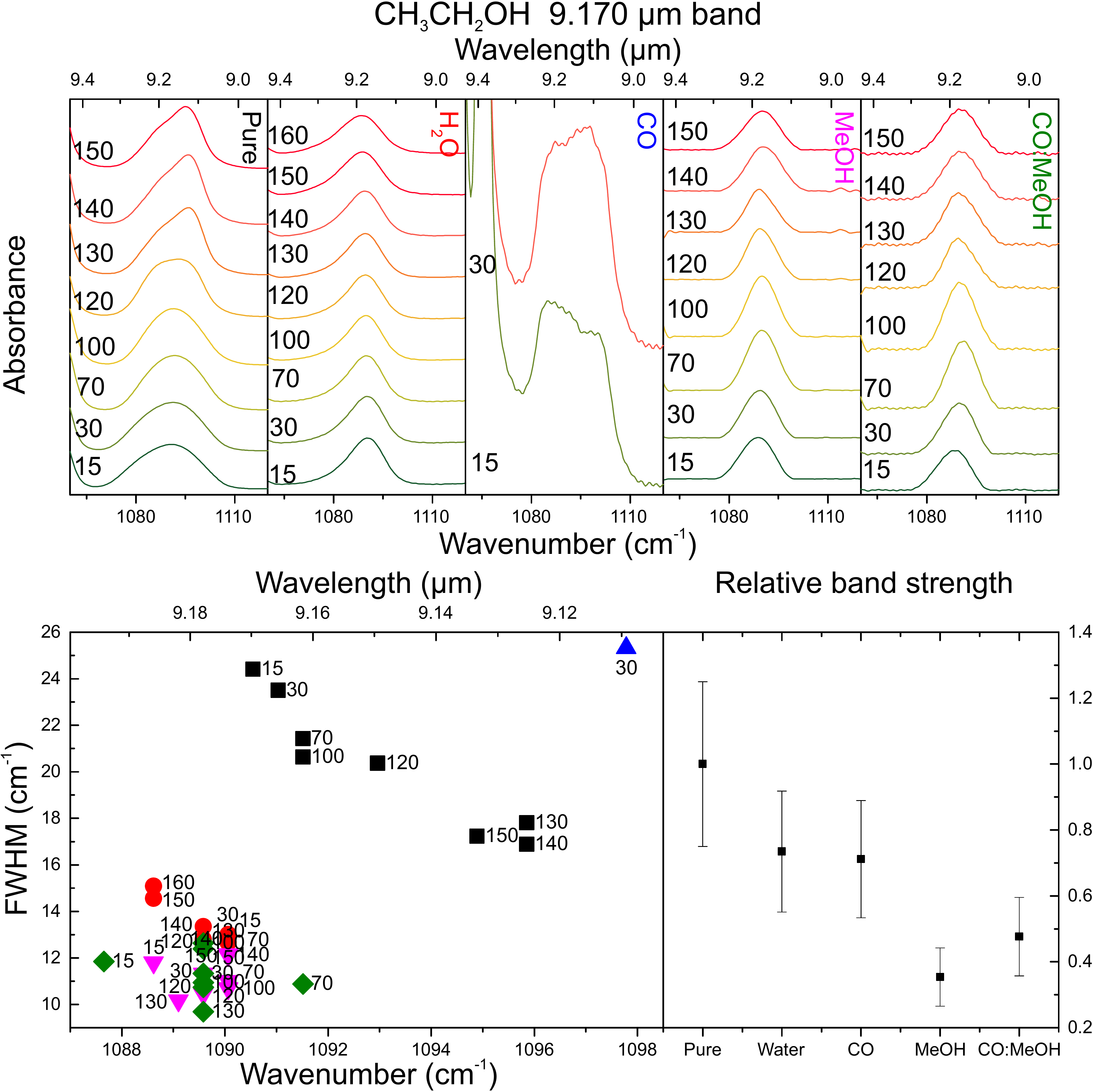}
\caption{Top: from left to right the ethanol 9.170 $\mu$m band pure (black) and in water (red), CO (blue), methanol (purple), and CO:CH$_{3}$OH (green) at various temperatures. Bottom left: peak position vs. FWHM plot, using the same colour coding. Bottom right: the relative band strength for the 9.170 $\mu$m band at 15~K in various matrices.}
\label{fig.E1095}
\end{figure*}

\begin{figure*}
\includegraphics[height=\hsize]{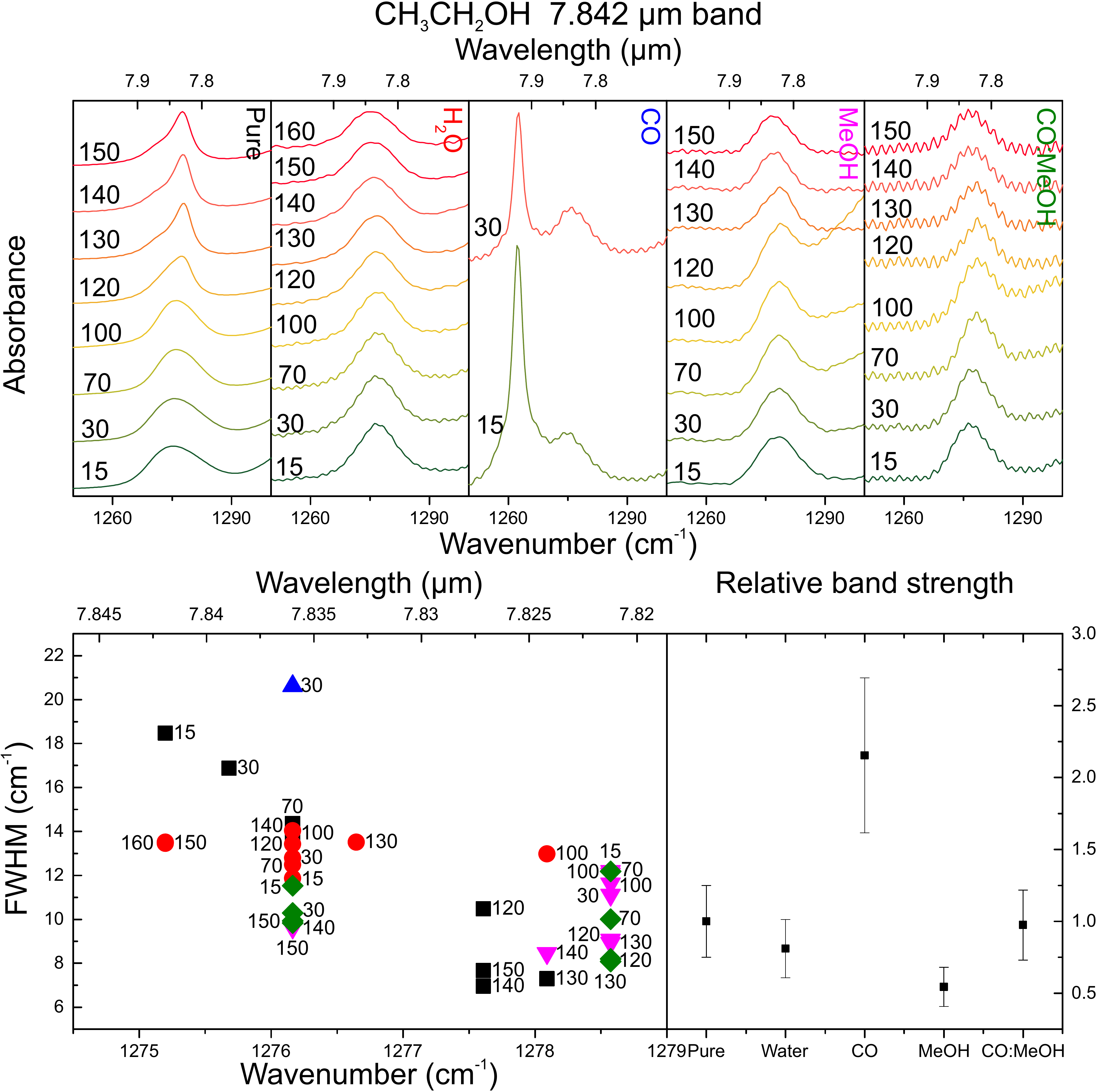}
\caption{Top: from left to right the ethanol 7.842 $\mu$m band pure (black) and in water (red), CO (blue), methanol (purple), and CO:CH$_{3}$OH (green) at various temperatures. Bottom left: peak position vs. FWHM plot, using the same colour coding. Bottom right: the relative band strength for the 7.842 $\mu$m band at 15~K in various matrices.}
\label{fig.E1278}
\end{figure*}

\begin{figure*}
\includegraphics[height=\hsize]{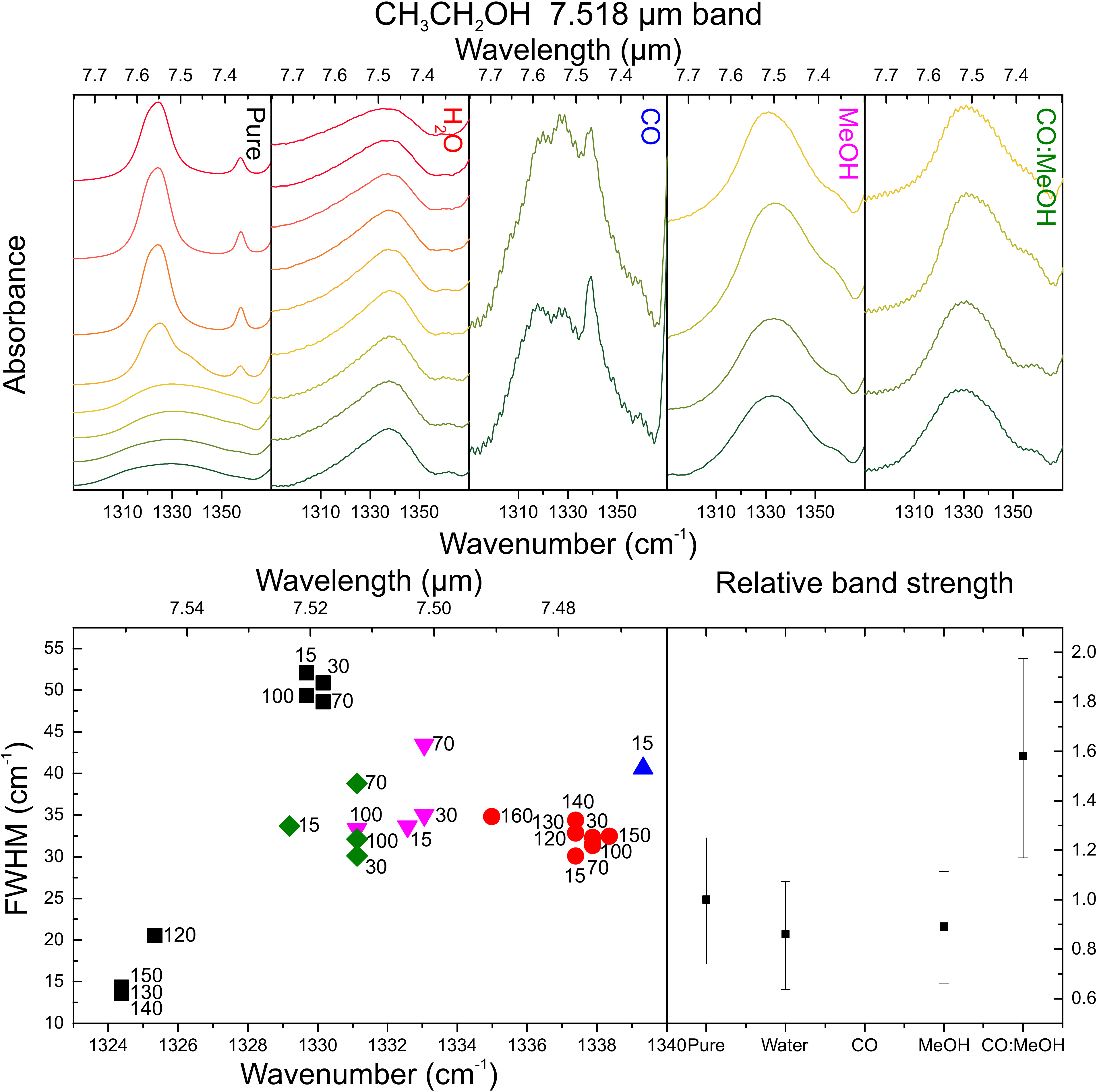}
\caption{Top: from left to right the ethanol 7.518 $\mu$m band pure (black) and in water (red), CO (blue), methanol (purple), and CO:CH$_{3}$OH (green) at various temperatures. Bottom left: peak position vs. FWHM plot, using the same colour coding. Bottom right: the relative band strength for the 7.518 $\mu$m band at 15~K in various matrices.}
\label{fig.E1324}
\end{figure*}

\subsection{Dimethyl ether}

\begin{figure*}
\includegraphics[height=\hsize]{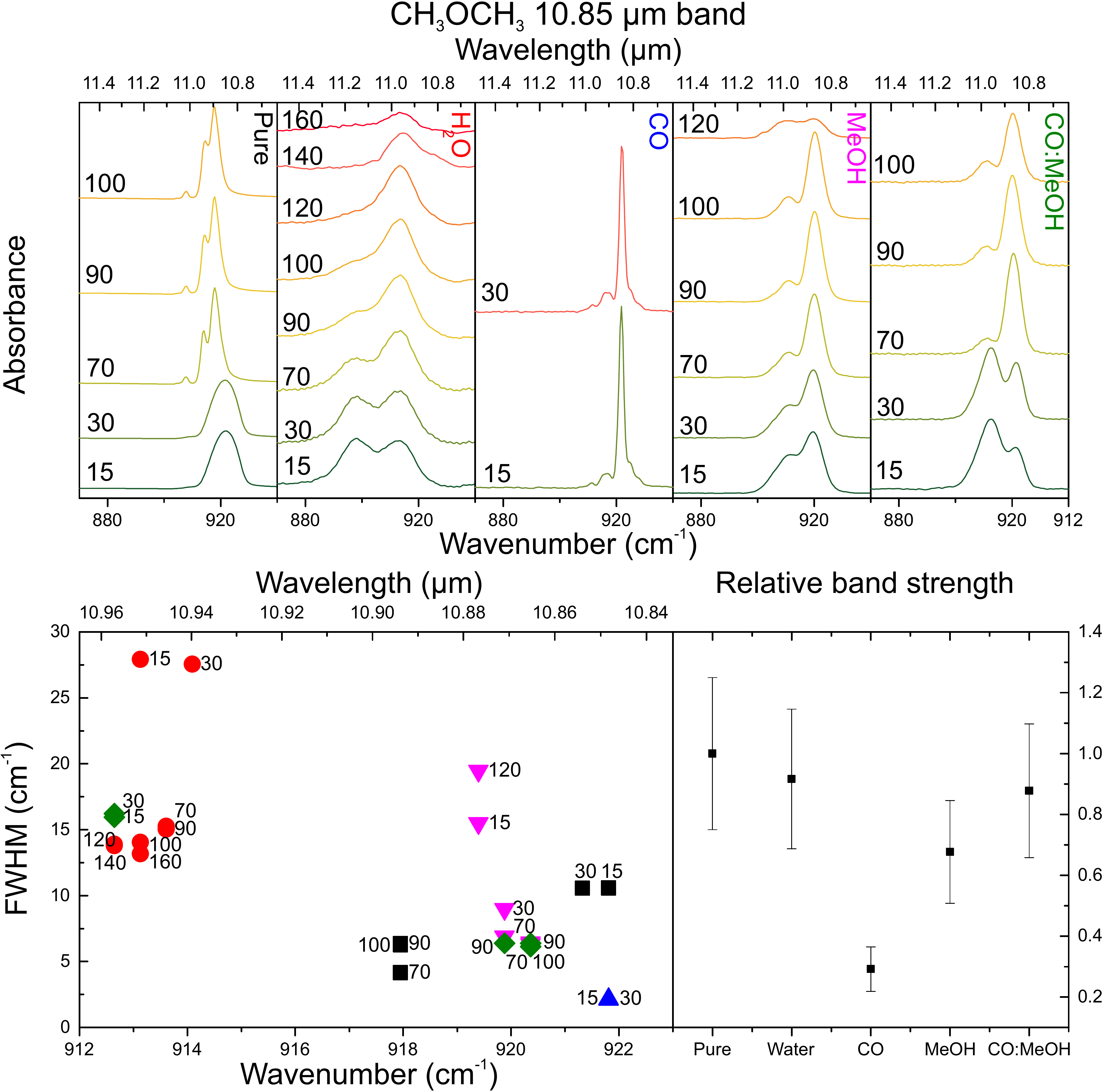}
\caption{Top: from left to right the dimethyl ether 10.85 $\mu$m band pure (black) and in water (red), CO (blue), methanol (purple), and CO:CH$_{3}$OH (green) at various temperatures. Bottom left: peak position vs. FWHM plot, using the same colour coding. Bottom right: the relative band strength for the 10.85 $\mu$m band at 15~K in various matrices.}
\label{fig.D922}
\end{figure*}

\begin{figure*}
\includegraphics[height=\hsize]{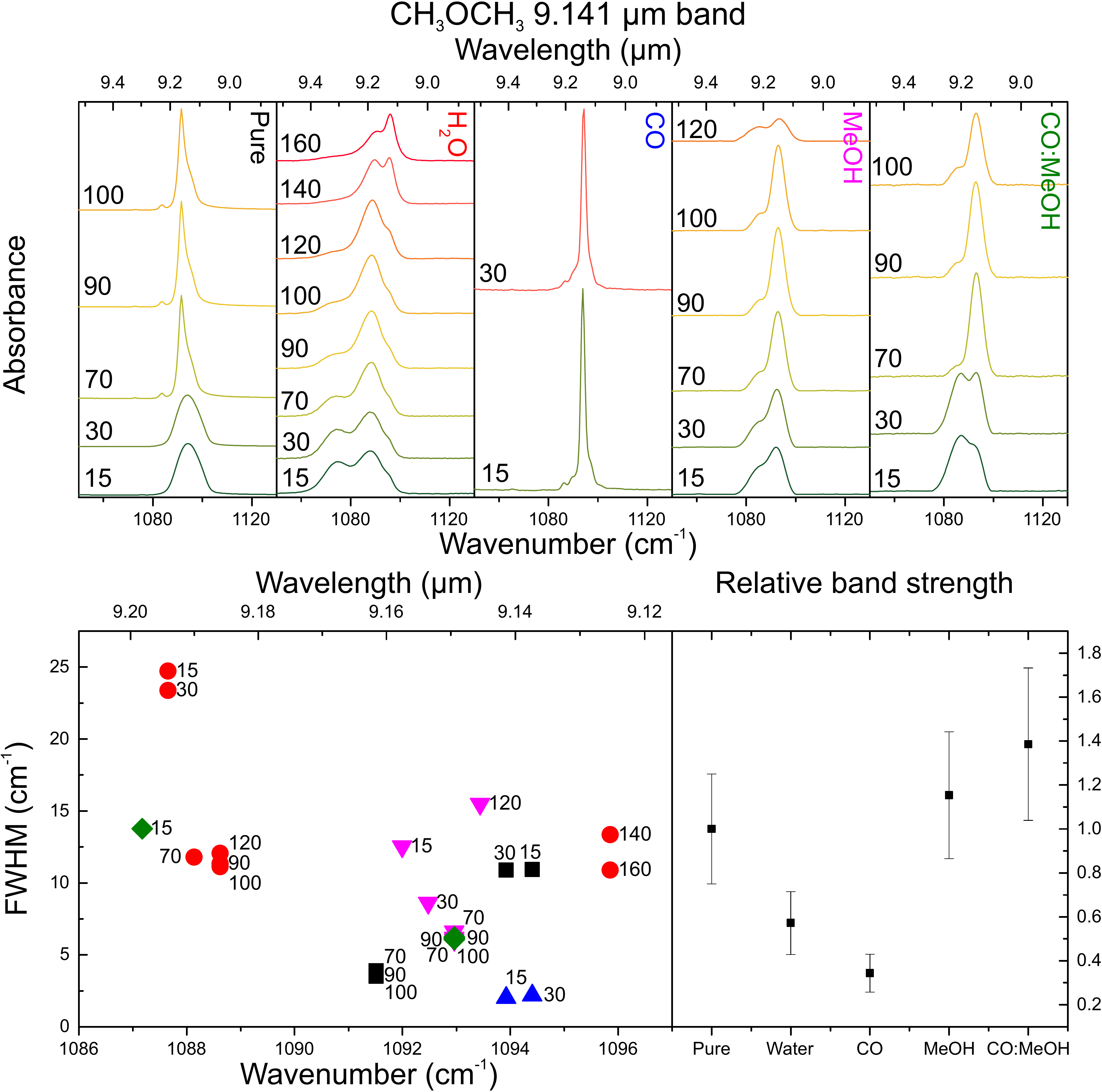}
\caption{Top: from left to right the dimethyl ether 9.141 $\mu$m band pure (black) and in water (red), CO (blue), methanol (purple), and CO:CH$_{3}$OH (green) at various temperatures. Bottom left: peak position vs. FWHM plot, using the same colour coding. Bottom right: the relative band strength for the 9.141 $\mu$m band at 15~K in various matrices.}
\label{fig.D1094}
\end{figure*}

\begin{figure*}
\includegraphics[height=\hsize]{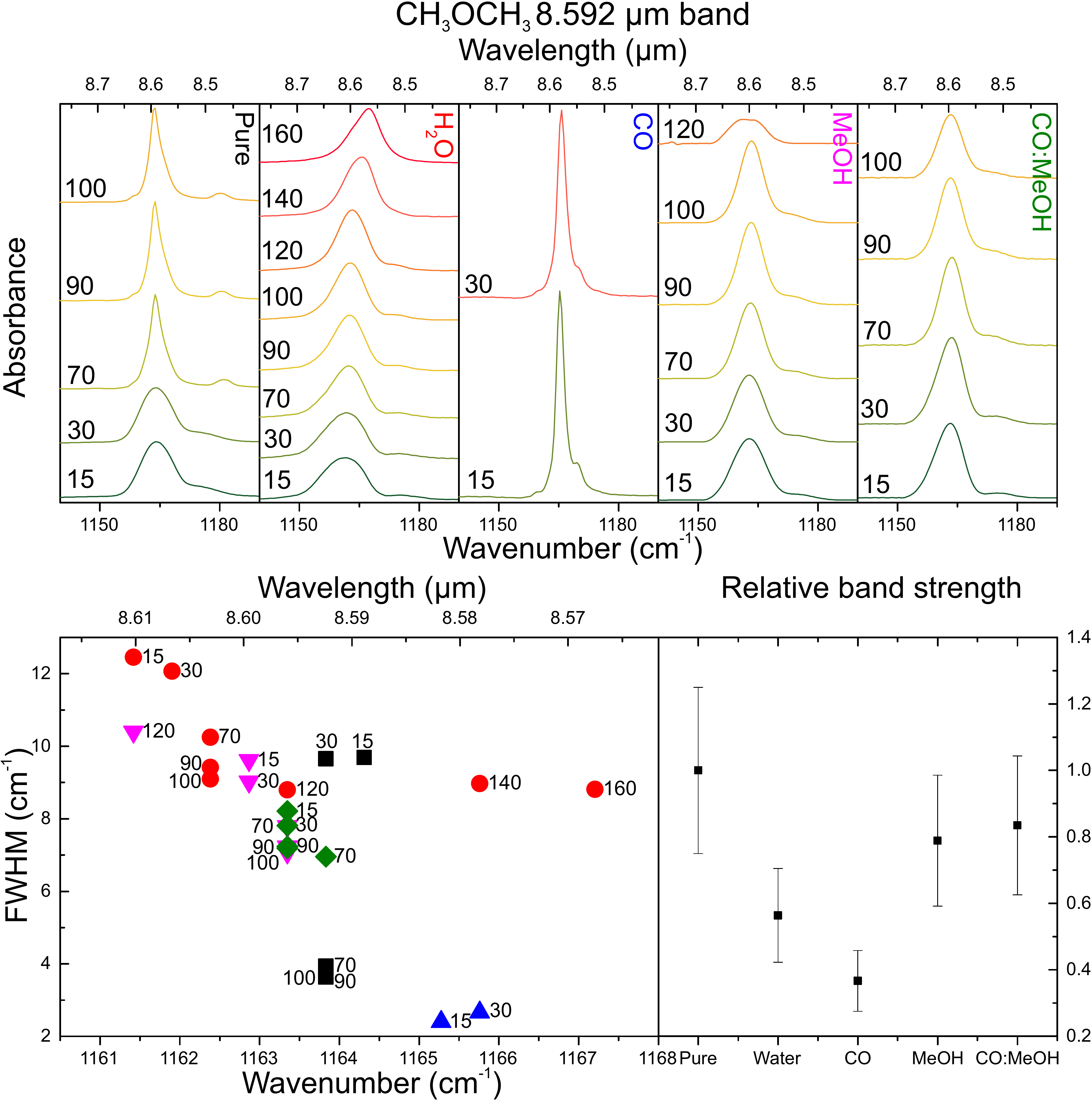}
\caption{Top: from left to right the dimethyl ether 8.592 $\mu$m band pure (black) and in water (red), CO (blue), methanol (purple), and CO:CH$_{3}$OH (green) at various temperatures. Bottom left: peak position vs. FWHM plot, using the same colour coding. Bottom right: the relative band strength for the 8.592 $\mu$m band at 15~K in various matrices.}
\label{fig.D1164}
\end{figure*}

%__________________________________________________________________
\onecolumn

\end{document}